\documentclass[sn-basic]{sn-jnl}

\usepackage{graphicx}%
\usepackage{multirow}%
\usepackage{amsmath,amssymb,amsfonts}%
\usepackage{hhline}
\usepackage{amsthm}%
\usepackage{mathrsfs}%
\usepackage[title]{appendix}%
\usepackage{xcolor}%
\usepackage{textcomp}%
\usepackage{manyfoot}%
\usepackage{booktabs}%
\usepackage{algorithm}%
\usepackage{algorithmicx}%
\usepackage{algpseudocode}%
\usepackage{listings}%
\usepackage[acronym, nohypertypes={acronym}]{glossaries}
\usepackage{subcaption}
\usepackage{lineno}
\usepackage[input-uncertainty-signs=\pm, separate-uncertainty = true, group-separator = {\,}, detect-weight=true, detect-family, print-unity-mantissa = false, parse-numbers=false]{siunitx}
\usepackage{hhline}
\usepackage{makecell}
\usepackage{ulem} 
\usepackage{pbox}
\newacronym{ll-SST}{ll-SST}{low-low satellite-to-satellite tracking}
\newacronym{LRI}{LRI}{laser range interferometer}
\newacronym{XHPS}{XHPS}{Extended High Performance Satellite Dynamics Simulator}
\newacronym{FE}{FE}{Finite Element}
\newacronym{SRP}{SRP}{solar radiation pressure}
\newacronym{GFR}{GFR}{gravity field recovery}
\newacronym{SGRS}{SGRS}{simplified gravitational reference sensor}
\newacronym{TM}{TM}{test mass}
\newacronym{ACME}{ACME}{Accelerometer Modeling Environment}
\newacronym{NGGM}{NGGM}{next generation gravimetry mission}
\newacronym{LISA}{LISA}{Laser Interferometer Space Antenna}
\newacronym{IGP}{IPG}{Institute for Gravitational Physcis}
\newacronym{QACC}{QACC}{Quantum Accelerometry}
\newacronym{ACS}{ACS}{attitude control system}
\newacronym{LoS}{LoS}{line of sight}
\newacronym{GaAs}{GaAs}{Galliumarsenid}
\newacronym{GRS}{GRS}{gravitational reference sensors}
\newacronym{RMS}{RMS}{root mean square}
\newacronym{ASD}{ASD}{amplitude spectral densities}

\begin{document}

\title[Article Title]{Evaluation of Deployable Solar Panels on GRACE-like Satellites by Closed-Loop Simulations}


\author*[1]{\fnm{Andreas} \sur{Leipner} \orcid{https://orcid.org/0000-0002-1317-9447}}\email{andreas.leipner@dlr.de}

\author[2]{\fnm{Alexey} \sur{Kupriyanov}  \orcid{https://orcid.org/0000-0002-0743-5889}}\email{kupriyanov@ife.uni-hannover.de}

\author[3,4]{\fnm{Arthur} \sur{Reis}  \orcid{https://orcid.org/0000-0002-6682-5457}}\email{arthur.reis@aei.mpg.de}

\author[2]{\fnm{Annike} \sur{Knabe}  \orcid{https://orcid.org/0000-0002-6603-8648}}\email{knabe@ife.uni-hannover.de}

\author[5]{\fnm{Manuel} \sur{Schilling}  \orcid{https://orcid.org/0000-0002-9677-0119}}\email{manuel.schilling@dlr.de}


\author[3,4]{\fnm{Vitali} \sur{Müller}  \orcid{https://orcid.org/0000-0002-2339-1353}}\email{vitali.mueller@aei.mpg.de}

\author[2,5]{\fnm{Matthias} \sur{Weigelt}  \orcid{https://orcid.org/0000-0001-9669-127X}}\email{matthias.weigelt@dlr.de}

\author[2]{\fnm{Jürgen} \sur{Müller}  \orcid{https://orcid.org/0000-0003-1247-9525}}\email{mueller@ife.uni-hannover.de}

\author[5]{\fnm{Meike} \sur{List}  \orcid{https://orcid.org/0000-0002-5268-5633}}\email{meike.list@dlr.de}


\affil[1]{\orgdiv{Institute for Satellite Geodesy and Inertial Sensing}, \orgname{German Aerospace Center (DLR)}, \orgaddress{\street{Am Fallturm 9}, \city{Bremen}, \postcode{28359}, \country{Germany}}}

\affil*[2]{\orgdiv{Institute of Geodesy}, \orgname{Leibniz University Hannover}, \orgaddress{\street{Schneiderberg 50}, \city{Hannover}, \postcode{30167}, \state{Lower Saxony}, \country{Germany}}}

\affil[3]{\orgdiv{Max Planck Institute for Gravitational Physics (IGP)}, \orgname{Albert Einstein Institute}, \orgaddress{\street{Callinstraße 38}, \city{Hannover}, \postcode{30167}, \state{Lower Saxony}, \country{Germany}}}

\affil[4]{\orgdiv{Institute for Gravitational Physics}, \orgname{Leibniz University Hannover}, \orgaddress{\street{Callinstraße 38}, \city{Hannover}, \postcode{30167}, \state{Lower Saxony}, \country{Germany}}}

\affil[5]{\orgdiv{Institute for Satellite Geodesy and Inertial Sensing}, \orgname{German Aerospace Center (DLR)}, \orgaddress{\street{Callinstraße 30B}, \city{Hannover}, \postcode{30167}, \state{Lower Saxony}, \country{Germany}}}

\abstract{
Future satellite gravimetry missions aim to exceed the performance of CHAMP, GOCE, GRACE, and GRACE-FO to meet growing scientific and user demands. These missions will incorporate advanced technologies, including novel inertial sensors such as optical and quantum accelerometers, high-precision inter-satellite laser ranging instruments, and potentially electric thrusters with micro-Newton thrust capabilities. However, increased power requirements for sensors and propulsion systems necessitate larger solar panel areas, while payload mass and launcher constraints impose significant design limitations. This study evaluates the implications of modified satellite shapes on \acrfull{GFR} by means of closed-loop simulation. Five satellite configurations were analyzed, including a standard shape and variations with single and double solar panels mounted on the top and bottom of the satellite body, each with corresponding finite element model and distinct moments of inertia. Detailed orbit simulations were carried out considering not only the non spherical static gravity field, but also time-variable background models of non gravitational forces. Performance of a modeled \acrfull{SGRS} equipped with an optical interferometer test mass displacement readout, was investigated. Also, since the air drag coefficient is a complex and non-trivial parameter that depends on multiple factors, it was varied from 2.25 for the standard shape to 4.5 for the double-panel setup. The time-variable gravity background models were excluded from the \acrshort{GFR} analysis, as their dominating influence would overshadow the effects of instrument performance. Evaluation of the retrieved gravity models was carried out in the spectral domain using the Degree RMS of spherical harmonic coefficient differences between the recovered and reference fields. The analysis showed that discrepancies between the modified and standard configurations primarily come from variations in the SGRS actuation noise, influenced by the satellite's cross-sectional area. Furthermore, the convergence of the residuals in the spectral domain in the \acrshort{GFR} results, when simulated orbits with different drag coefficients were applied to the double-panel configuration, confirmed the dominating role of the \acrshort{SGRS} performance in the retrieved gravity field.
}

\keywords{future satellite gravimetry missions, finite element modeling, satellite shapes, optical accelerometry, gravity field recovery, closed-loop simulation}

\maketitle

\section{Introduction}\label{sec_intro}

Over the past two decades, satellite gravimetry missions have yielded distinctive datasets fundamental to a wide range of scientific and practical applications \citep{Wiese.2022}. Three key missions, CHAMP \citep{torge.2023}, GOCE \citep{Flechtner.2021}, and GRACE \citep{Chen.2022}, alongside the currently operational GRACE-FO mission \citep{Peidou.2022}, have contributed gravity field data products at varying spatial and temporal resolutions.

The data produced by these missions have been applied extensively in geophysical and environmental research. Examples include the analysis of glacier mass balance in Antarctica and Greenland \citep{Otosaka.2023}, estimation of Glacial Isostatic Adjustment \citep{Kang.2022}, terrestrial water storage assessments \citep{Humphrey.2023}, and monitoring of gravitational variations originating from deep Earth processes \citep{Chen.2022}. Furthermore, satellite gravimetry has proven useful in natural hazard assessments (e.g., floods and droughts) \citep{sun.2017} and investigations into Earth’s polar motion and length-of-day trends \citep{Zotov.2022}.

Satellite gravimetry employs two primary measurement principles. The first approach, exemplified by GOCE, involves single-satellite gravity gradiometry, using onboard gradiometers to measure gravity gradients \citep{Brockmann.2014}. The second approach, pioneered by the GRACE and GRACE-FO missions, involves the measurement of inter-satellite distance variations in a \acrlong{ll-SST} (\acrshort{ll-SST}) configuration, using microwave ranging \citep{Tapley.2019}. GRACE-FO further demonstrated the \acrlong{LRI} (\acrshort{LRI}) as a pathfinder instrument, anticipated to be a standard feature in future gravimetry missions \citep{muller.2022}.

Additionally, the concept of drag-compensation, first implemented in GOCE, utilized an ion propulsion system to mitigate orbit decay and prevent accelerometer saturation, albeit with introduced thruster noise \citep{Canuto.2018}. This system probably will also feature in ESA’s forthcoming \acrlong{NGGM} (\acrshort{NGGM}), a paired-satellite mission in an inclined orbit designed for enhanced spatial-temporal resolution with five-day repeat cycles \citep{Daras.2023}.

Despite these advancements, the spatial-temporal resolution offered by GRACE-like missions remains insufficient for numerous applications. Consequently, research has focused on augmenting mission accuracy, which can be categorized into two main approaches: enhancing inertial sensors, particularly accelerometers, and exploring alternative satellite formations.

The electrostatic accelerometers used in previous missions, such as GRACE(-FO) and GOCE, exhibit low-frequency drift that limits their precision \citep{vanCamp.2021}. However, the stringent requirements of the \acrlong{LISA} (\acrshort{LISA}) gravitational wave detector have driven significant advancements in inertial sensing technologies, demonstrated with the \acrshort{LISA}-Pathfinder mission \citep{Armano.2021}. These advancements have inspired a range of proposals for new electrostatic and optical accelerometers suitable for Earth observation \citep{Davila.2022, Weber.2022}. Additional studies have investigated the use of cold atom interferometers, both standalone and hybridized, for gravimetric sensing applications \citep{Zahzam.2022, HosseiniArani.2024}.

Further studies have evaluated the potential of multiple satellite pairs \citep{Heller-Kaikov.2023}, novel tracking concepts such as high-low satellite-to-satellite tracking \citep{Pail.2019}, configurations permitting cross-track measurements \citep{Kupriyanov.2023b}, and small satellite constellations for monitoring sub-daily mass changes \citep{Pfaffenzeller.2023}.

Given the increased power demands anticipated for future \acrshort{ll-SST} missions incorporating drag-compensated platforms and advanced inertial sensors \citep{Dionisio.2018}, the use of larger solar arrays is under consideration. However, constraints on payload mass and volume within the launch vehicle fairing limit satellite size \citep{Haagmans.2020}, motivating the development of deployable solar panel systems.

A key research question driving this study is whether the modified satellite shape, necessitated by the integration of novel inertial sensors, drag compensation system and non-optimized solar illumination could compromise the quality of scientific data collected and reducing mission lifetime in future gravimetry missions. The need for highly accurate gravity field measurements is critical for advancing understanding of mass transport processes within the Earth system \citep{Pail.2015, Wiese.2022}. The altered satellite geometry may introduce asymmetries in mass distribution and surface area exposed to atmosphere, thereby increasing susceptibility to drag and thermal radiation effects. These perturbations could induce additional noise in the inter-satellite ranging measurements and degrade the performance of onboard accelerometers, potentially undermining the anticipated improvements from advanced inertial sensors.

This study evaluates modified satellite designs with respect to \acrlong{GFR} (\acrshort{GFR}). The paper is structured as follows: Chapter \ref{sec_Challenges_of_shapes} discusses the challenges associated with future gravimetry missions, focusing on the implications of modified satellite shapes due to deployable solar panels. Chapter \ref{sec_software_overview} provides an overview of the software used. Section \ref{sec_xhps} details the satellite dynamics simulation framework. In section \ref{sec_ACC_model}, an accelerometer model featuring optical readout for test mass (TM) displacement is presented. Section \ref{sec_GFR_assumptions} explains the functional and stochastic modeling in \acrshort{GFR} simulations. Major error sources and assumptions that took place in the simulation procedure are listed in the section \ref{error_evaluation}. Chapter \ref{sec_Res} discusses the orbit simulation and GFR results. The paper ends with Chapter \ref{ch: Discussion and conclusions} where discussion of the results is presented.

\section{Challenges of future gravimetry missions and modeling of the modified satellite shapes}\label{sec_Challenges_of_shapes}

Proposals for next-generation gravimetry missions include the deployment of satellite pairs in constellation configurations, similar to current \acrshort{ll-SST} missions \citep{Haagmans.2020}. However, several mission parameters are anticipated to differ in these successor missions. To enhance gravity signal resolution, future missions may operate at lower orbital altitudes, which will amplify the gravity signal but also increase the influence of atmosphere disturbance forces. Mission lifetimes are projected to span at least one full 11-year solar cycle to ensure comprehensive temporal coverage \citep{Dionisio.2018}, during which the satellites will experience the full variation in \acrlong{SRP} (\acrshort{SRP}) effects. 

An ion thruster system is planned to enable enhanced attitude and orbit control, including precise drag compensation \citep{Haagmans.2020}. This system is estimated to require approximately $\qty{1}{\kW}$ of power \citep{Dionisio.2018}; however, this demand could increase with the inclusion of advanced electrostatic accelerometers or inertial sensors equipped with optical \acrlong{TM} (\acrshort{TM}) readouts on future missions. Additionally, the non-sun-synchronous orbits proposed for these missions will result in varying solar illumination conditions, impacting the solar panels' power generation. This variability necessitates careful consideration in detailed pre-mission analysis to ensure adequate power supply.

For reference, the currently operating GRACE-FO mission, equipped with gallium arsenide solar arrays, achieves a power production of approximately $\qty{1.37}{\kW}$ \citep{Kornfeld.2019}. However, this may prove insufficient for future missions, given the anticipated higher power demands.

The mission design requires that both satellites be launched and deployed in orbit simultaneously to establish the necessary paired configuration. Critical constraints include the geometrical limitations of the launch vehicle fairing and the total allowable payload mass, which future gravimetry missions must accommodate to ensure mission viability \citep{Dionisio.2018, Haagmans.2020}.

\begin{figure}[!htbp]
	\centering		
    	\begin{subfigure}[b]{0.45\textwidth}
            \centering
    		\includegraphics[width=0.7\linewidth]{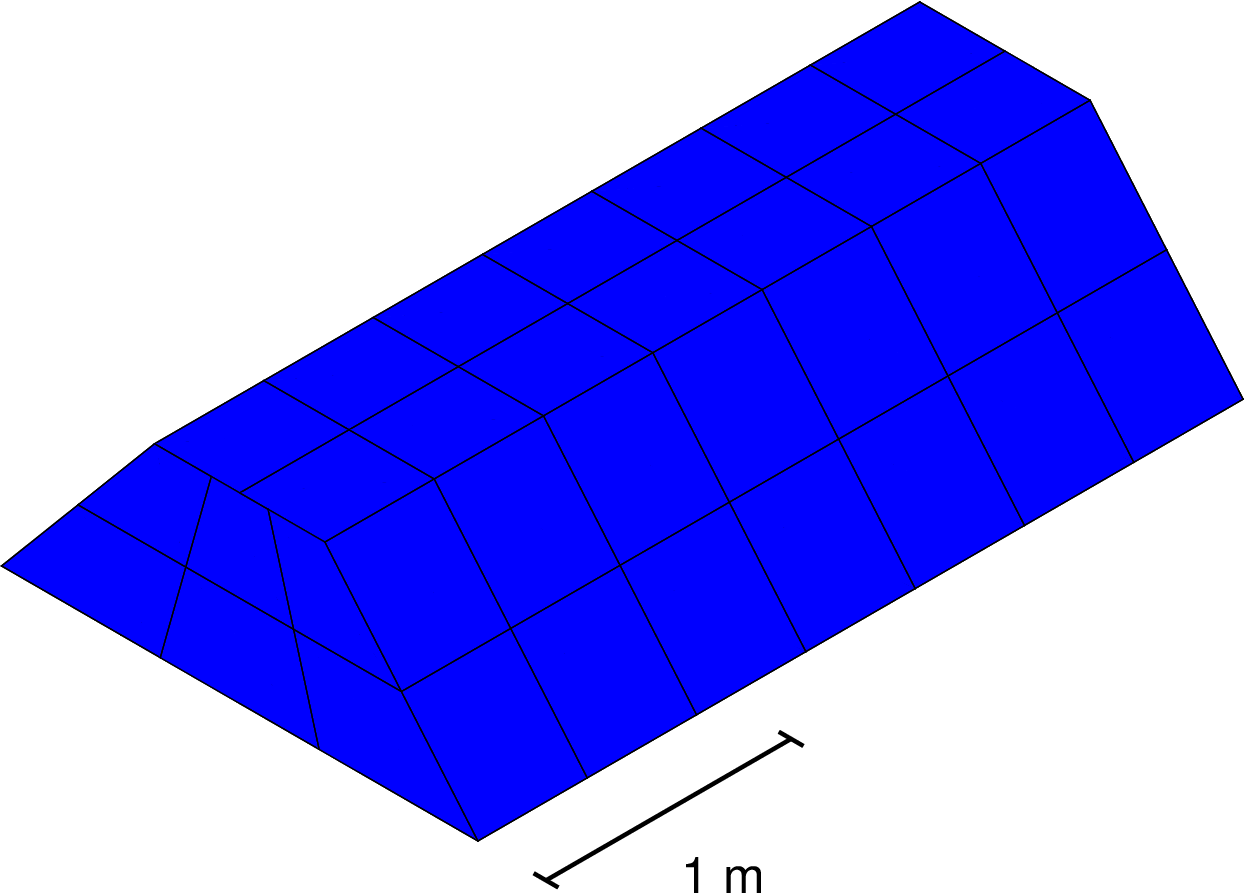} 
    		\caption{GRACE-FO standard shape}
    		\label{fig: fem_GRACE}
    	\end{subfigure}\\ 
    	\begin{subfigure}[b]{0.45\textwidth}
    		\includegraphics[width=\linewidth]{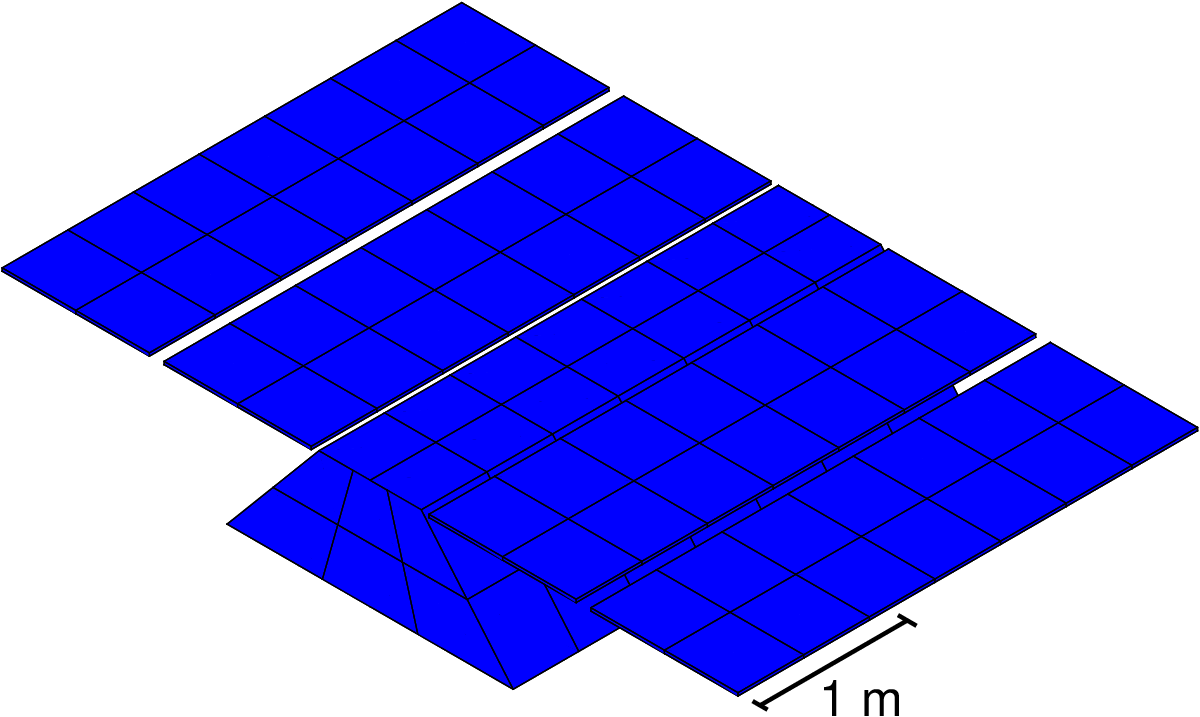}
    		\caption{GRACE-like shape (double solar panel top)}
    		\label{fig: fem_GRACE_top}
    	\end{subfigure}
    \hspace{0.2cm}
        \begin{subfigure}[b]{0.45\textwidth}
    		\includegraphics[width=\linewidth]{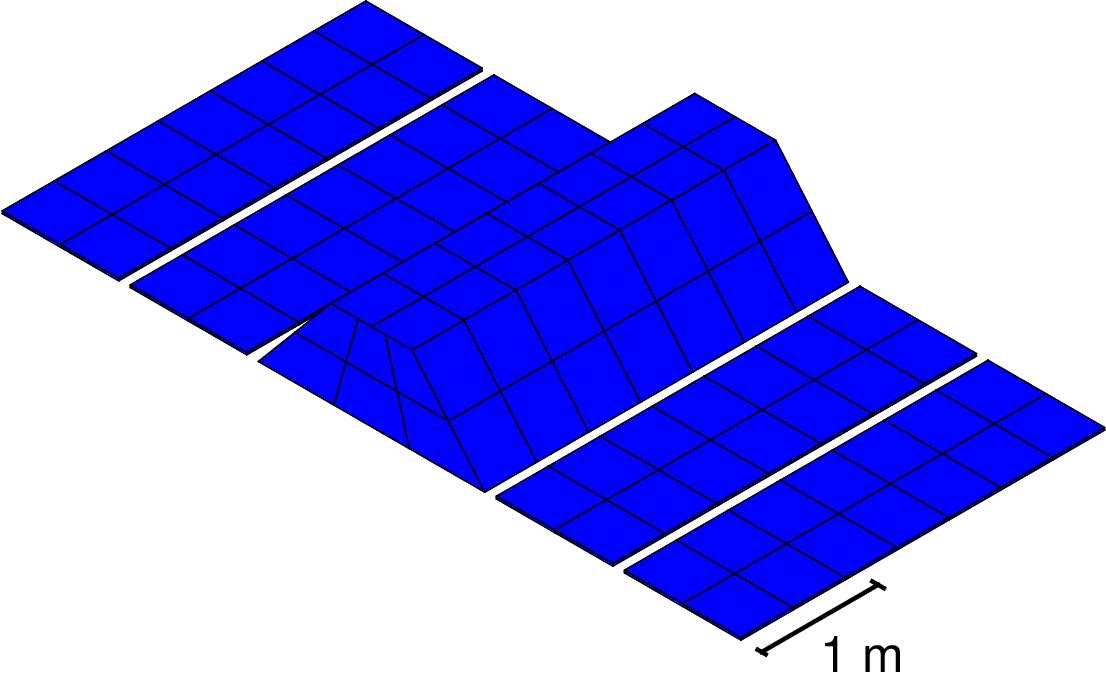}
    		\caption{GRACE-like shape (double solar panel bottom)}
    		\label{fig: fem_GRACE_bottom}
    	\end{subfigure}
         \begin{subfigure}[b]{0.45\textwidth}
    		\includegraphics[width=\linewidth]{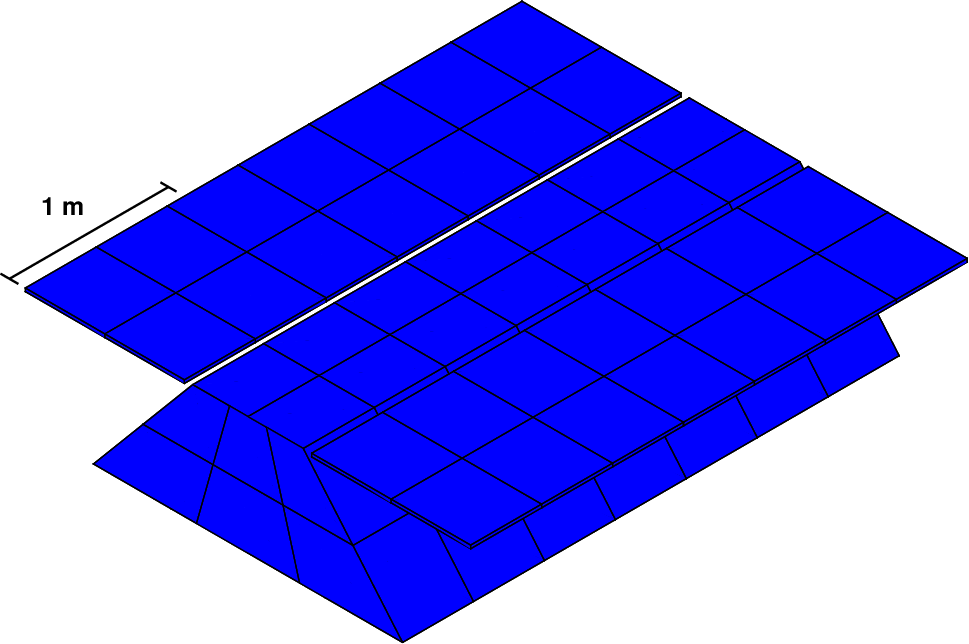}
    		\caption{GRACE-like shape (single solar panel top)}
    		\label{fig: fem_GRACE_single_top}
    	\end{subfigure}
    \hspace{0.2cm}
        \begin{subfigure}[b]{0.45\textwidth}
    		\includegraphics[width=\linewidth]{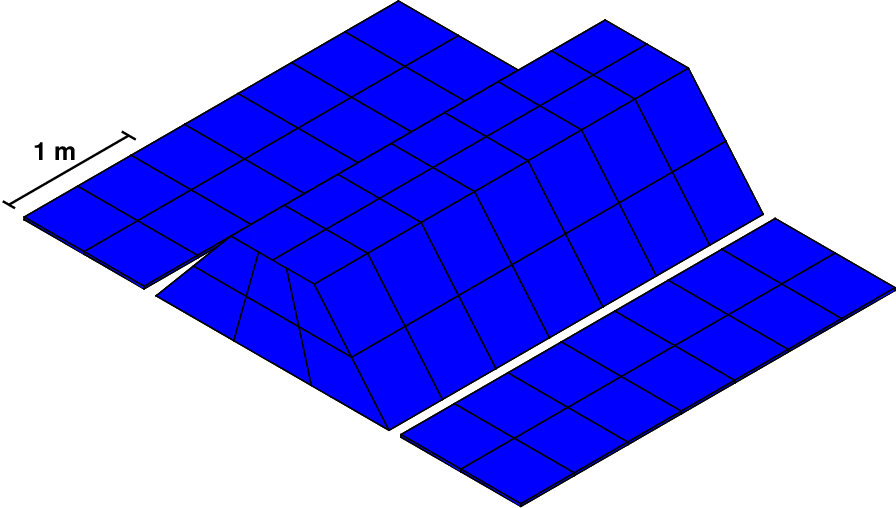}
    		\caption{GRACE-like shape (single solar panel bottom)}
    		\label{fig: fem_GRACE_single_bottom}
    	\end{subfigure}	
	\caption{\acrshort{FE} models of the studied satellite shapes.}
	\label{fig: fem_models}
\end{figure}

To address the anticipated power demands of future gravimetry satellites, deployable solar panels present a feasible solution. These panels can remain folded during launch, minimizing the satellite’s surface dimensions within the fairing constraints. Once deployed, however, the expanded surface area will increase susceptibility to atmospheric drag and \acrshort{SRP}. As a result, the effects of this increased drag and \acrshort{SRP} must be carefully modeled and accounted for in the gravity field recovery process to maintain data accuracy.

To evaluate the implications of these expanded surface areas, this study employed \acrlong{FE} (\acrshort{FE}) modeling to simulate different satellite shapes inside \acrlong{XHPS} (\acrshort{XHPS}). By comparing shapes with deployable panels in varied positions, such as top-mounted and bottom-mounted orientations, the simulated models allowed for detailed analysis of how these designs impact \acrshort{GFR}.

The \acrshort{FE} models of satellite configurations evaluated in this study are presented in Fig. \ref{fig: fem_models}. For comparison, the standard GRACE-FO satellite body, represented by a trapezoidal prism, is shown in Fig. \ref{fig: fem_GRACE}, with dimensions and mass consistent with the actual GRACE-FO satellite ($\qty{1.942}{\m}\times\qty{3.123}{\m}\times\qty{0.72}{\m}$, $\qty{601.214}{\kg}$ \citep{NASA.2002}). The surface properties (reflectivity, absorption and emission \citep{NASA.2002}) are assigned to the corresponding finite elements to maintain consistency with the actual satellite configuration. Additionally, four modified satellite designs incorporating deployable single or double solar panels, mounted either on the top (Fig. \ref{fig: fem_GRACE_top}/\ref{fig: fem_GRACE_single_top}) or bottom (Fig. \ref{fig: fem_GRACE_bottom}/\ref{fig: fem_GRACE_single_bottom}) of the satellite body, are evaluated. These configurations could potentially increase power generation by approximately $\qty{1}{\kW}$ for the single solar panel or $\qty{2}{\kW}$ for the double solar panels, meeting the enhanced energy requirements for future mission instruments and control systems.

An additional consideration is the potential obstruction of the star tracker cameras' field-of-view due to various solar panel configurations. While this effect was not addressed in this study, it remains essential to account for these implications in future mission design.

\section{Overview of the closed-loop simulation procedure and software parts}\label{sec_software_overview}
\begin{figure*}[!hbtp]
\centering
\noindent\includegraphics[width=\textwidth]{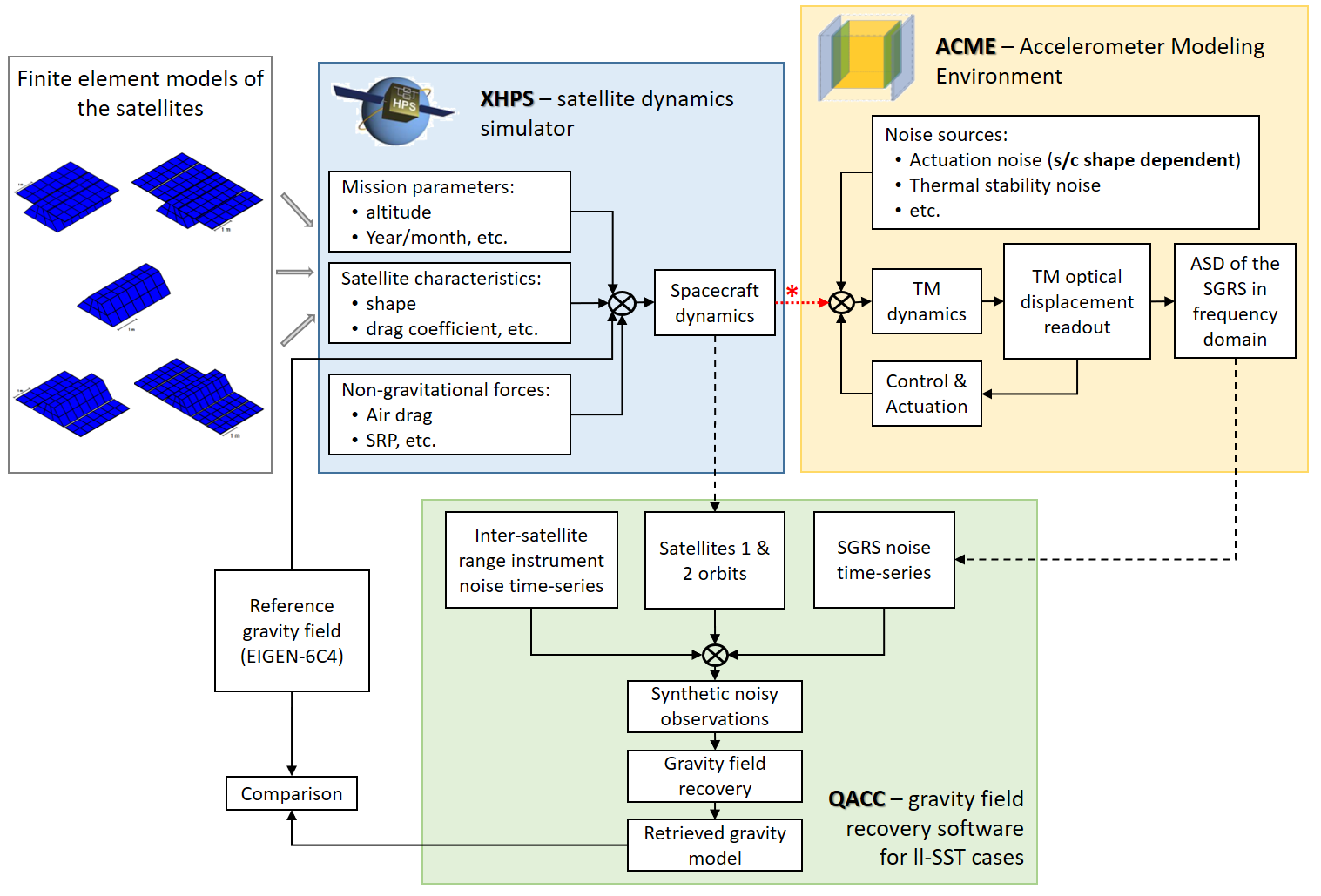}
\caption{Block diagram of the closed-loop simulation procedure. * simplifications of orbit dynamics has been considered. ASD - Amplitude Spectral Density; QACC - Quantum Accelerometry (gravity field recovery tool for ll-SST)}
\label{fig: closed_loop_procedure}
\end{figure*}

The simulation methodology applied in this study is depicted in Fig. \ref{fig: closed_loop_procedure}. It consists of multiple interconnected modules, each representing a distinct software tool that can function independently or, as in this instance, as part of an integrated simulation and data analysis workflow. The combination of the \acrlong{XHPS} (XHPS) and \acrlong{ACME} (ACME) modules facilitates the generation of scientific datasets for a prospective \acrshort{ll-SST} mission, while the Quantum Accelerometry (QACC) software module is employed to derive the corresponding gravity field solutions.

Orbital disturbance forces are strongly influenced by the satellite’s complex geometry, even when assuming a homogeneous distribution of optical properties across its external surfaces \citep{List.2015}. To accurately capture the effects of these disturbance forces for the \acrshort{FE} models from Fig. \ref{fig: fem_models}, corresponding lookup tables were generated through a preprocessing algorithm to facilitate subsequent use in the satellite dynamics simulator.

The orbital dynamics were simulated using the \acrshort{XHPS} software developed by ZARM/DLR \citep{woske.2016}. \acrshort{XHPS} enables precise definition of key mission parameters, including the initial position and velocity of the satellite(s), as well as their orientation. The software incorporates a \acrshort{FE} model of the satellite(s) to allow detailed modeling of disturbance forces based on satellite geometry. Additionally, the simulation incorporates non-gravitational forces along with a provided reference gravity field model into the orbit integrator.

The multi-degree-of-freedom accelerometer was modeled at the Max Planck \acrlong{IGP} (IGP) using a Matlab/Simulink-based tool named \acrlong{ACME} (\acrshort{ACME}) \citep{Kupriyanov.2023a}. The model parameters adhered to the specifications outlined in the proposed \acrlong{SGRS} (\acrshort{SGRS}) framework \citep{Davila.2022}. Operating concurrently with \acrshort{XHPS}, \acrshort{ACME} generated the accelerometer noise budget, represented as amplitude spectral densities in the frequency domain.

\acrshort{GFR} simulations were executed on the Leibniz University IT Services (LUIS) cluster \citep{LUIS_cluster.2024} using the \acrshort{QACC} toolbox, developed at the Institute of Geodesy (IfE) \citep{Wu.2016}. In this setup, synthetic noisy observations are generated based on satellite positions from \acrshort{XHPS}, incorporating various modeled error sources. Specifically, the stochastic component encompasses all error contributors, including inter-satellite range instrument noise time-series and, sourced from \acrshort{ACME}, \acrshort{SGRS} noise time-series. Subsequently, gravity field recovery is performed, and the retrieved gravity model is compared to the reference model (EIGEN-6C4), with results evaluated based on the residuals. The following sections provide detailed descriptions of each module within the simulation procedure.

\subsection{Satellite dynamics simulator}\label{sec_xhps}

The satellite dynamics simulator's (\acrshort{XHPS}) primary purpose is to generate data that can be used in the \acrshort{GFR} process. It is designed to accurately model the satellite's behavior by taking into account various disturbance forces such as non-spherical gravitational fields, atmospheric drag and \acrshort{SRP}. Additionally, the simulator incorporates the geometry of the satellite(s) into its calculations on the basis of \acrshort{FE} models. In that way, it provides a detailed and realistic dataset that helps understand the impact of satellite shapes on \acrshort{GFR}, ultimately aiding in the refinement of satellite geodesy techniques and improving the precision of geophysical measurements. 

\paragraph{Earth's gravitational field}

The modelling of gravitational fields in the simulator is based on spherically harmonics, which provide a mathematical framework for representing the Earth's gravitational field \citep{Wertz.1980}:
\begin{align}
\label{eq:SH}
V(r,\phi,\lambda) =& \frac{\mu}{r} \sum^{\infty}_{l=0} \left( \frac{a_e}{r} \right)^{l}
    \sum^{l}_{m=0} P_{l,m} \left( \sin{\phi} \right) \cdot 
    \left[ C_{l,m} \cos{m \lambda} + S_{l,m} \sin{m \lambda} \right]\text{,}
\end{align}
where $\mu$ is the product of the universal constant of gravitation $G$ and the mass of the Earth $M$, $a_e$ is the semi-major axis of the Earth's reference ellipsoid, $r,\phi,\lambda$ are the satellite distance, latitude and longitude, respectively in a body-fixed coordinate system, $C_{l,m}$ and $S_{l,m}$ are the spherical harmonic coefficients of degree $l$ and order $m$ and $P_{l,m}$ are the Associated Legendre Functions of degree $l$ and order $m$. Each term in the series corresponds to a different frequency component, capturing variations in the gravitational potential, including irregularities such as the equatorial bulge and local anomalies. This allows for a description of the field and determination of the gravitational acceleration on the satellite.

\paragraph{Atmospheric drag}

 Atmospheric drag is known to be one of the most disturbing effects at the appropriate orbit height for gravimetry missions \citep{Rievers.2019}. The simulator is accounting for variations in the Earth's atmosphere at mission's orbit height, as well as the disturbance acceleration due to the geometry of the satellite. The force is calculated via the \acrshort{FE} model approach. 

The atmospheric drag force $d\Vec{f}_{\textrm{drag}}$ acting on one of the \acrshort{FE} model elements with the outward normal vector $\Vec{n}$ and surface area $dA$ can be calculated by \citep{Wertz.1980}

\begin{equation}
    \label{eq:drag}
    d\Vec{f}_{\textrm{drag}} = -\frac{1}{2} c_D \rho v^2 (\Vec{n} \cdot \Vec{v}) \Vec{v} dA\text{,}
\end{equation}
where $ \Vec{v}$ is the unit vector in the direction of the satellites velocity relative to the atmosphere around it, $\rho$ is the atmospheric density, $c_D$ is the drag coefficient and $v$ is the norm of the relative velocity vector. The total atmospheric drag force can be calculated by summing up all forces acting on each \acrshort{FE} model element. The atmospheric density is calculated during the simulation at each time step and current satellite location. This ensures a high resolution in time and space of the drag disturbance force.

\paragraph{Solar radiation pressure}
Precise orbital calculations and long-term mission planning, particularly for satellites with large surface areas, must account for the forces exerted on the satellite by solar radiation. The simulator accounts for solar disturbance acceleration by calculating it using the \acrshort{FE} model approach and assuming parallel Sun rays striking the satellite.

The \acrshort{SRP} force $d\Vec{f}_{\textrm{SRP}}$ acting on one of the \acrshort{FE} model elements with the outward normal vector $\Vec{n}$ and surface area $dA$ can be calculated as \citep{Wertz.1980}
\begin{align}
    \label{eq:SRP}
    d\Vec{f}_{\textrm{SRP}} =& -P \left[ \left( 1 - c_s \right) \Vec{s} +  2 \left( c_s \cos{\theta} + \frac{1}{3} c_d \right) \Vec{n} \right]\cos{\theta}dA\text{,}
\end{align}
where $P$ is the solar flux at the satellites position, $\Vec{s}$ is the unit vector from the satellite to the sun and $\theta$ is the angle between the \acrshort{FE} model element normal vector $\Vec{n}$ and the vector $\Vec{s}$. The coefficents $c_s$ and $c_d$ are for specular and diffuse reflection. Like for the atmospheric drag force, the total \acrshort{SRP} force can be calculated by summing over all forces acting on each \acrshort{FE} model element. Earth's shadow is implemented enterly geometric, so a disc is shadowing the satellite.

These forces are highly dependent on the satellites' shape, as the effective area, force coefficients, and force directions are influenced by the satellites' geometry. For more detailed explaination on the selected satellite shapes see section \ref{sec_Challenges_of_shapes}. 

The simulator implements the equations for drag and solar radiation pressure within a preprocessing algorithm that generates lookup tables to manage the complex computations required for precise force determination. Initially, this algorithm calculates the drag force and solar radiation pressure force for a specified range of incidence angles. The precomputed values are then stored in lookup tables, which the simulator references during runtime to efficiently and accurately determine disturbance forces at each time step.

In the simulations, the gravitational field was modelled using EIGEN-6C4 \citep{Forste.2014}, truncated to degree and order 90. For third-body perturbations from the Sun and Moon, the DE430 ephemeris \citep{2014IPNPR.196C...1F} was applied. Atmospheric drag forces were calculated using the NRLMSISE-00 density model \citep{Picone.2002}. A drag coefficient of $c_D=2.25$ was assigned to the GRACE-FO shape \citep{Woske.2018}, consistent with typical values for convex-shaped satellite \citep{Montenbruck.2002}. For the modified satellite configurations (see Fig. \ref{fig: fem_models}), a drag coefficient of $ c_D = 3.0 $ was assumed to account for the added surface area from the solar panels. Subsequently, for the shape of the top mounted double solar panels values of $ c_D = 2.65 $ and $ c_D = 4.5 $ (doubled value of GRACE-FO shape) were considered to analyse the effect of $c_D$ values. These drag coefficients were taken as a potential minimum and maximum of possible values for such advanced satellite bodies, based on \citep{Moe.2005} and \citep{Hassa.2013}. The solar flux for the \acrshort{SRP} force was set to a constant value of $W = 1368 \frac{\textrm{J}}{\textrm{sm}^2}$ throughout the simulations. The pertubation models are listed in Table \ref{tab:Simulation Models}.

\begin{table}[!htbp]
    \centering
    \caption{Background models considered in orbital simulations. [1] - \citep{Forste.2014}; [2] - \citep{2014IPNPR.196C...1F}; [3] - \citep{Picone.2002}}
    \begin{tabular}{l|l|l}
        \hline
        \textbf{Pertubation} & \textbf{Model} & \textbf{Remark}\\
        \Xhline{4\arrayrulewidth}
        Static gravity field & EIGEN-6C4 [1]& considered till d/o 90\\
        \hline
        Ephimeris model & DE430 [2]& -\\
        \hline
        \makecell[l]{Third bodies: \\ - Sun \\ - Moon }& - & -\\
        \hline
        Non-gravitational Forces: & - & -\\
        \hline
        Air drag & \makecell[l]{Density model: \\ \hspace{0.2cm}NRLMSISE-00 [3]}& \makecell[l]{- Effective area reevaluated each time step}\\
        \hline
        \acrlong{SRP} & \makecell[l]{\acrshort{SRP} calculation: \\  \hspace{0.2cm}parallel sun rays\\ Constant solar flux: \\\hspace{0.2cm}$W = 1368 \frac{\textrm{J}}{\textrm{sm}^2}$} & \makecell[l]{- Eclipse calculation: geometric\\ \hspace{0.2cm}cylindrical eclipse\\ - Ilumination condition reevaluated each \\ \hspace{0.2cm}time step}\\
        \hline
    \end{tabular}
    
    \label{tab:Simulation Models}
\end{table}

The initial conditions for all satellite pairs and simulations are listed in Table \ref{tab:Simulation Initial Values}. These initial parameters, including the start date of January 1, 2008, and an attitude control system mode of "\acrfull{LoS} Pointing" via a PID controller, were kept constant for all simulations to isolate the effects of the satellite shapes. The values of position and velocity are provided by the GROOPS software \citep{MayerGurr.2021}, developed by the Institute of Geodesy at Graz University of Technology, based on the GRACE orbits.

\begin{table}[!htbp]
    \centering
    \caption{Initial conditions for all satellites with initial date, position, velocity and \acrfull{ACS}. Position and velocity of the satellites are provided by GROOPS \citep{MayerGurr.2021}}
    \begin{tabular}{lrrrr} 
        \toprule 
        \textbf{Satellite} & \textbf{Position} [$\textrm{m}$] & \textbf{Velocity} [$\frac{\textrm{m}}{\textrm{s}}$]& \textbf{Date}& \textbf{\acrshort{ACS}} \\ 
        \midrule[2pt]
        Satellite A & \makecell[r]{x: \hspace{0.22cm}$1.04106912\cdot10^6$\\y: \hspace{0.22cm}$3.37233841 \cdot 10^6$\\z: $-5.87039994\cdot 10^6$} & \makecell[r]{x: $1.68961\cdot10^3$\\y: $6.30209\cdot10^3$\\z: $3.93698\cdot10^3$}& \makecell[r]{2008-01-08 \\00:00:00} & \acrshort{LoS}\\
        \midrule
        Satellite B & \makecell[r]{x: \hspace{0.22cm}$1.09381033\cdot10^6$\\y: \hspace{0.22cm}$3.57174758\cdot10^6$\\z: $-5.74121787\cdot10^6$} & \makecell[r]{x: $1.64626\cdot10^3$\\y: $6.16464\cdot10^3$\\z: $4.16732\cdot10^3$}& \makecell[r]{2008-01-08 \\00:00:00} & \acrshort{LoS}\\
        \bottomrule
    \end{tabular}
    
    \label{tab:Simulation Initial Values}
\end{table}

The satellite’s mass is set to $\qty{601.214}{\kg}$, with its inertia tensor defined as specified in Table \ref{tab:Satellite properties} \citep{Wen.2019}. For the modified configurations featuring single solar panels, an additional mass of $\qty{7.8}{\kg}$ is included, calculated based on the density of $\qty{1.25}{\frac{\kg}{m^2}}$ for the triple-junction \acrlong{GaAs} (\acrshort{GaAs}) GRACE-FO solar panels \citep{satsearch.03.07.2024}. For configurations with double solar panels, an additional mass of $\qty{15.6}{\kg}$ is accounted for. The inertia tensor for the modified satellite shapes is computed using the parallel axis theorem, applied to the inertia tensor of the GRACE-like shape to account for the added mass and the altered mass distribution. The resulting values are listed in Table \ref{tab:Satellite properties}.

\begin{table}[!htbp]
    \centering
    \caption{Satellite properties for simulation}
    \begin{tabular}{lrr} 
        \toprule
        \textbf{Satellite} & \textbf{Mass} [$\textrm{kg}$] & \textbf{Inertia Tensor} [$\textrm{kg}\textrm{m}^2$]\\ 
        \midrule[2pt]
        GRACE-FO & 601.214 &$ \left( \begin{matrix} 110.49 & -1.02 & 0.35 \\ -1.02 & 580.67 & 0.04 \\ 0.35 & 0.04 & 649.69 \end{matrix} \right) $ \\
        \midrule
        \makecell[l]{GRACE-like\\Single Solar Panel Top} & 609.014 &$ \left( \begin{matrix} 120.76 & -1.02 & 0.35 \\ -1.02 & 588.50 & 0.04 \\ 0.35 & 0.04 & 664.81 \end{matrix} \right) $ \\
        \midrule
        \makecell[l]{GRACE-like\\Single Solar Panel Bottom} & 609.014 &$ \left( \begin{matrix} 132.79 & -1.02 & 0.35 \\ -1.02 & 587.56 & 0.04 \\ 0.35 & 0.04 & 677.78 \end{matrix} \right) $ \\
        \midrule
        \makecell[l]{GRACE-like\\Double Solar Panels Top} & 616.814 &$ \left( \begin{matrix} 155.92 & -1.02 & 0.35 \\ -1.02 & 596.33 & 0.04 \\ 0.35 & 0.04 & 704.84 \end{matrix} \right) $ \\
        \midrule
        \makecell[l]{GRACE-like\\Double Solar Panels Bottom} & 616.814 &$ \left( \begin{matrix} 191.53 & -1.02 & 0.35 \\ -1.02 & 594.46 & 0.04 \\ 0.35 & 0.04 & 742.32 \end{matrix} \right) $ \\
        \bottomrule
    \end{tabular}
    
    \label{tab:Satellite properties}
\end{table}

\subsection{Simplified GRS optical accelerometer modeling in ACME}\label{sec_ACC_model}
This study employs an optical accelerometer, referred to as the \acrfull{SGRS}, as the inertial sensor. In the context of future satellite gravimetry missions, the impact of modified satellite shapes on gravity field recovery is, in principle, independent of the specific type of accelerometer used. However, optical inertial sensors offer several advantages over electrostatic, cold atom interferometry (CAI)-based, or hybrid instruments. Optical accelerometers, also known as \acrfull{GRS}, were successfully operated during the LISA Pathfinder mission. As a result, they benefit from a higher technology readiness level due to their proven flight heritage compared to CAI instruments. Furthermore, they offer significant performance enhancements over state-of-the-art electrostatic accelerometers, such as SuperSTAR. These improvements include advanced test mass displacement readout techniques utilizing optical interferometry, a cubic test mass design enabling highly sensitive multi-axis measurements, and a wireless charge management system, among other innovations \citep{Kupriyanov.2023a, Kupriyanov.2023b}. These features make optical accelerometers a compelling choice for improving gravity field recovery.

The \acrshort{SGRS} model used in this study is primarily based on the parameters proposed in \citep{Davila.2022}, incorporating a laser interferometer for test mass displacement readout. It is implemented within the \acrshort{ACME} simulation environment. However, compared to the \acrshort{SGRS} described in \citep{Davila.2022}, the thermal stability in the low-frequency domain is still under evaluation. Its impact is assumed to be more pronounced than reported in \citep{Davila.2022}, with thermal noise characteristics similar to those observed in the MicroSTAR electrostatic accelerometer \citep{christophe.2018, dalin.2020}.

Additionally, for actuation noise—related to the stability of the reference voltage applied to the actuation electrodes—a reference voltage of \qty{10}{\V} with a stability of \qty{0.6}{ppm/\sqrt{Hz}} was assumed, following the specifications in \citep{Mance.2012, Halloin.2013}. All other accelerometer noise sources were adopted as detailed in \citep{Kupriyanov.2023a}. The simulations considered a freely levitating cubic gold-platinum test mass with \qty{40}{\mm} sides and a \qty{1}{\mm} gap between the test mass and the surrounding electrode housing.

Given that this study examines various satellite shapes, the actuation noise level of the inertial sensor deserves more detailed attention. Non-gravitational forces influence the \acrshort{TM} through the coupling between the satellite and the \acrshort{TM} \citep{Josselin.1999}. Consequently, the actuation noise level is expected to vary depending on the cross-sectional area of the satellite, as this directly affects the magnitude of the coupling forces.

Simplifications were introduced in the modeling of non-gravitational forces to estimate the actuation noise of the \acrshort{SGRS} (denoted with an asterisk in Fig. \ref{fig: closed_loop_procedure}). Specifically, the system was treated as a single degree of freedom in the along-track direction, with atmospheric drag considered the dominant non-gravitational perturbation and the sole non-conservative force accounted for in \acrshort{ACME}. Under these assumptions, the resultant acceleration of a satellite with mass $M$ in the along-track direction is dependent on the local atmospheric density $\rho$, the satellite velocity \( v \), the cross-sectional area $A_{SC}$, and the drag coefficient $c_D$.

\begin{equation}
    a_{\text{non-grav}}= \frac{1}{2} \frac{\rho v^2 c_D A_{SC}}{M},
\end{equation}

which is a simplified version of equation \ref{eq:drag}. The density can be determined without running the complex models that are built-in XHPS by using a simpler atmospheric model such as Harris-Priester, where the density is a function of altitude $\rho = \rho(h)$ \citep{Cappellari.1976}.

 Assuming a circular orbit with altitude $h$ over the Earth as measured from the surface, the velocity along track is just \citep{Wertz.1980}

\begin{equation}
    v = \sqrt{\frac{G M_\oplus}{(R_\oplus + h)}}.
\end{equation}

The coulombic force that levitates a \acrshort{TM} of mass $m$, that forms a capacitor with an electrode of area $A_{el}$ and gap $d$, which is maintained in a potential difference $V$ is \citep{Josselin.1999}

\begin{equation}
    F = ma_\text{actuation} = \frac{1}{2}\frac{\epsilon_0 A_{el}}{d^2}V^2.
    \label{eq: Coulomb Force}
\end{equation}

Where $\epsilon_0$ is the dielectric constant of vacuum. Since the actuation of the electrode keeps the TM centered in its housing, the non-gravitational accelerations perceived by the satellite are thus coupled to the TM:
\begin{equation}
    |a_\text{non-grav}| = |a_\text{actuation}|
\end{equation}

inserting these equations into equation \ref{eq: Coulomb Force} we have the value of voltage required to constantly re-center the TM:

\begin{equation}
V = \sqrt{\frac{d^2}{\epsilon_0 A_{el}} \frac{m}{M} c_D A_{SC} \rho(h) \left(\frac{G M_\oplus}{R_\oplus + h}\right)}.
\end{equation}

Therefore, even in a back-of-the-envelope scenario, where everything is ideal but the voltage reference stability, we expect an associated actuation noise dependent not only on the accelerometer design but also on the satellite and orbit characteristics. 

\begin{figure}[!hbtp]
	\centering	 
    	\begin{subfigure}[b]{0.45\textwidth}
    		\includegraphics[width=\linewidth]{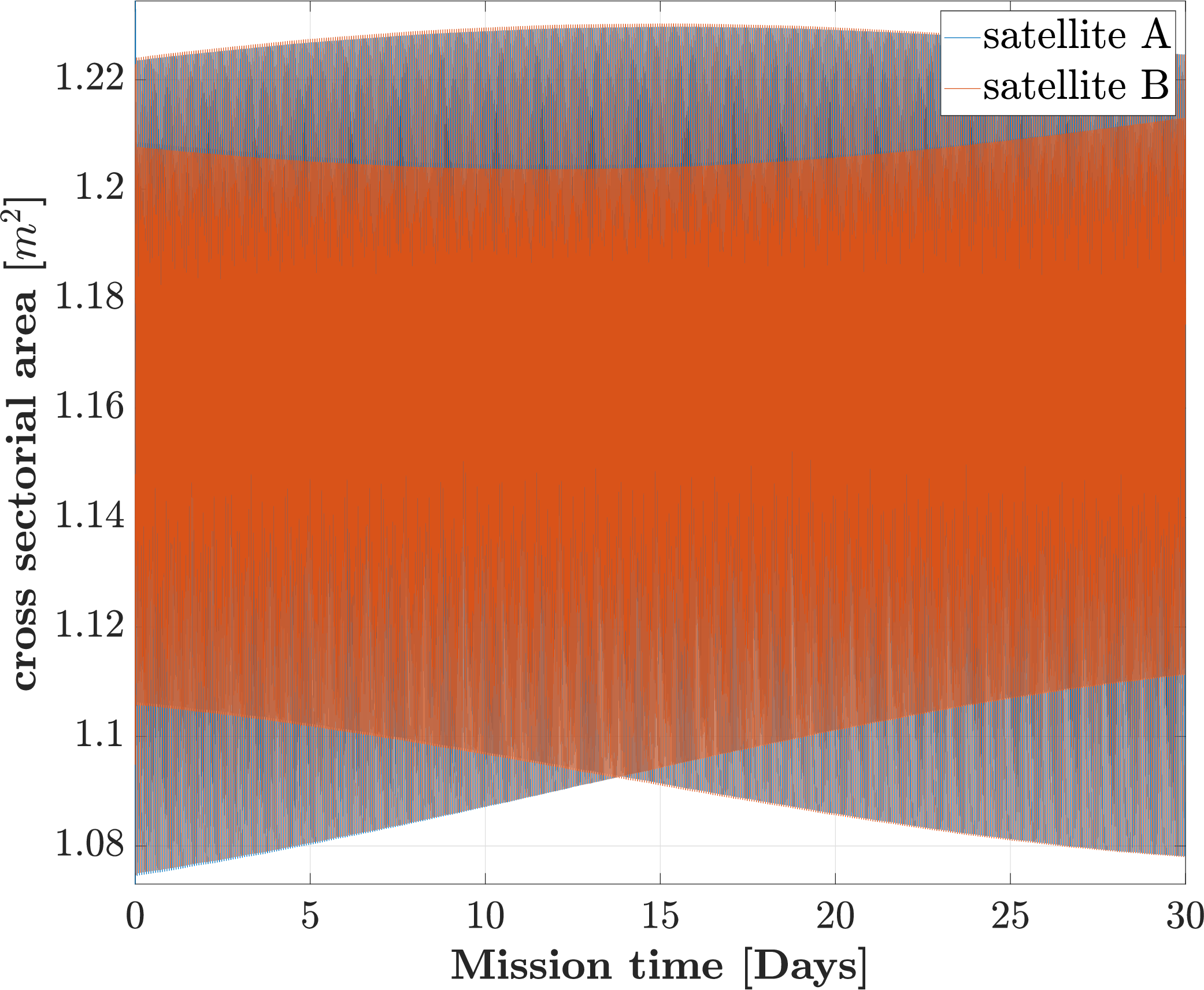}
    		\caption{standard GRACE-FO shape}
    		\label{fig: cross_sec_Simple}
    	\end{subfigure}
    \hspace{0.2cm}
        \begin{subfigure}[b]{0.45\textwidth}
    		\includegraphics[width=\linewidth]{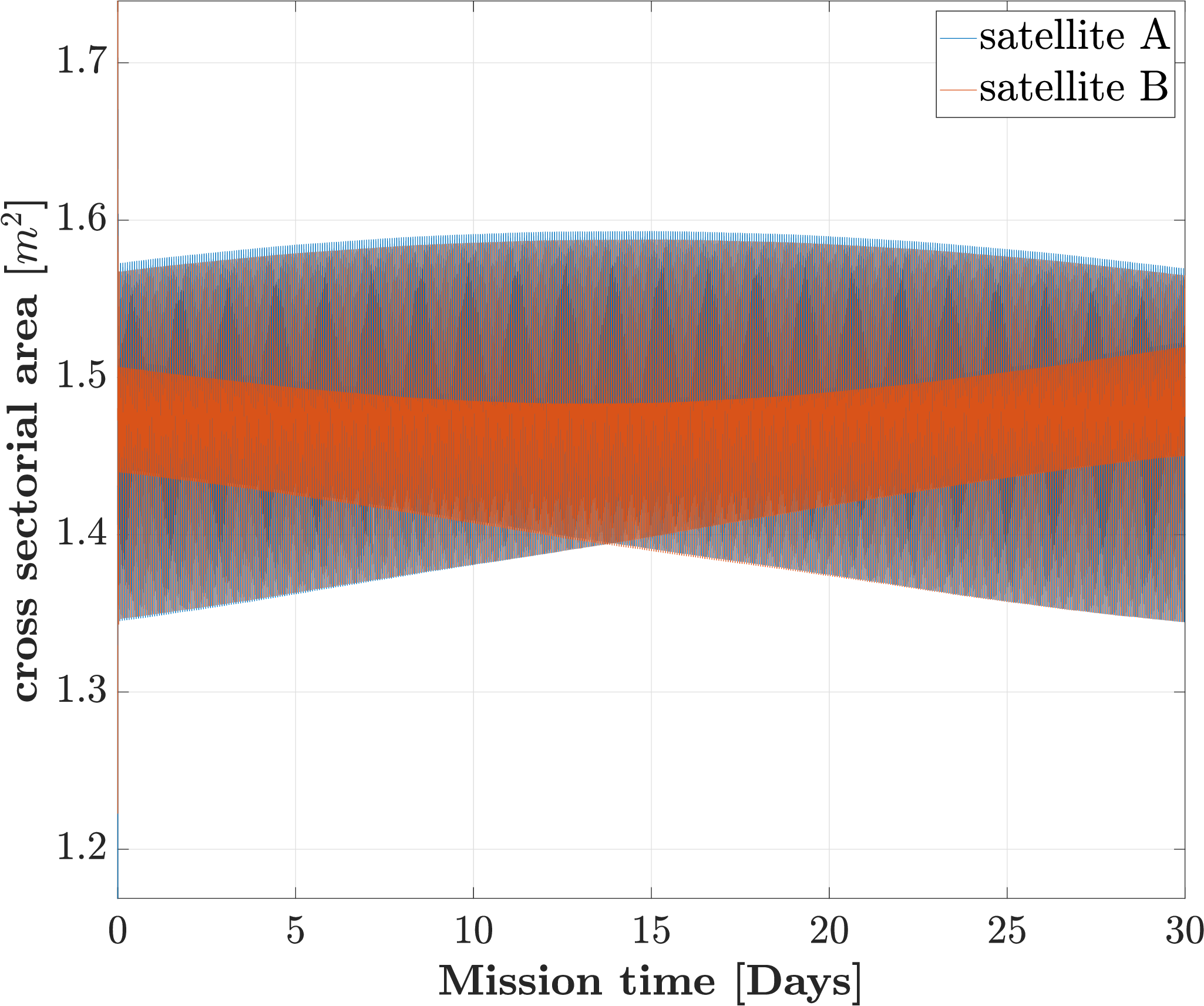}
    		\caption{GRACE-like shape (double solar panel)}
    		\label{fig: cross_sec_double}
    	\end{subfigure}
         \begin{subfigure}[b]{0.45\textwidth}
    		\includegraphics[width=\linewidth]{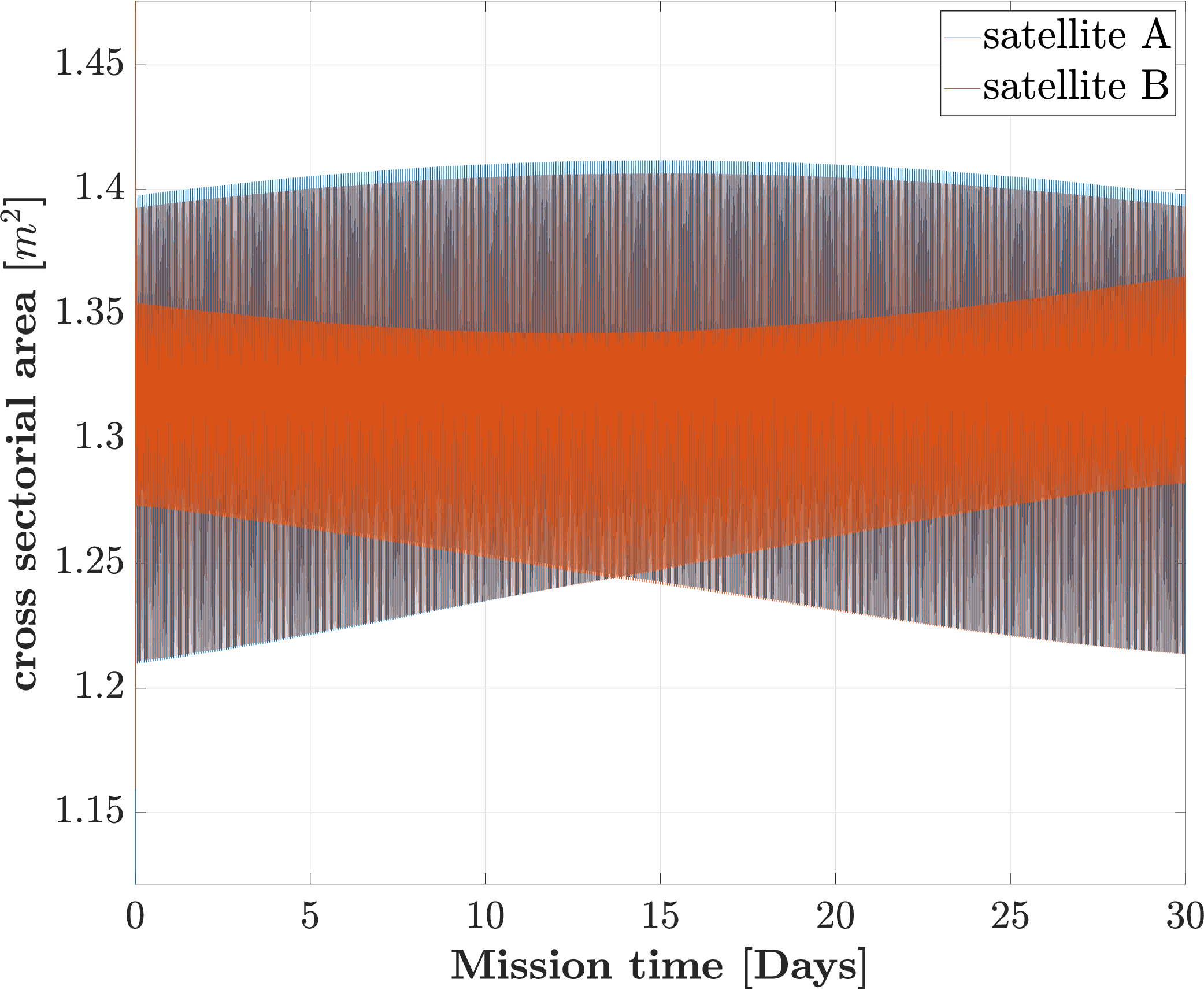}
    		\caption{GRACE-like shape (single solar panel)}
    		\label{fig: cross_sec_Single}
    	\end{subfigure}
	\caption{Cross-sectorial areas of the investigated satellite shapes during simulation.}
	\label{fig: cross_sectorial_area}
\end{figure}

From the orbital simulations carried out in XHPS, a value for the monthly averaged cross-sectorial area of \qty{1.16}{\m^2} was derived for the GRACE-FO standard shape, \qty{1.32}{\m^2} for the GRACE-like case with a single panel and \qty{1.47}{\m^2} for the GRACE-like case with double solar array (see Fig. \ref{fig: cross_sectorial_area}).

\begin{figure}[!htbp]
    \centering
    \noindent\includegraphics[width=\linewidth]{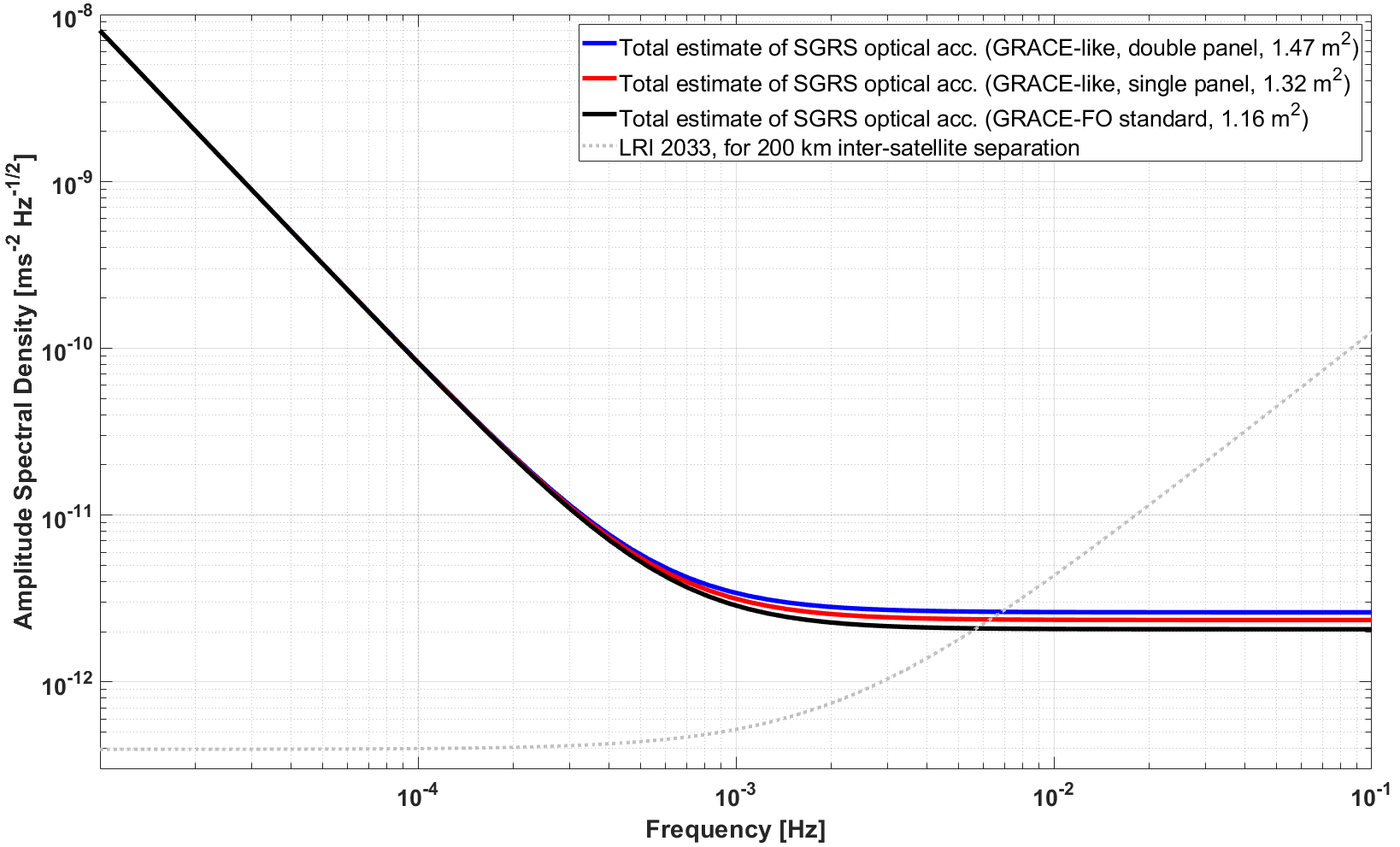}
    \caption{Amplitude spectral density of the total noise budget of the modeled SGRS optical accelerometers for the `standard' GRACE-FO shape (black curve) and modified ones (blue and red curves) and the anticipated measurement error of the inter-satellite \acrshort{LRI} in 2033 for the satellites separated by \qty{200}{\km} (grey dotted line).}
    \label{fig: ASD_SGRS_comparison}
\end{figure}

Fig. \ref{fig: ASD_SGRS_comparison} shows the \acrfull{ASD} of the modeled inertial sensors and inter-satellite \acrshort{LRI} that were used in the gravity field recovery simulations. The total noise budget in terms of \acrshort{ASD} of the modeled \acrshort{SGRS} for the `standard' (black curve), GRACE-like with single panel (red curve) and GRACE-like with double solar array (blue curve) is shown. The difference between the curves above \qty{1}{\mHz} is due to the different level of the actuation noise. Also, an amplitude spectral density of the measurement error of the inter-satellite laser range interferometer \citep{Kupriyanov.2023a} (grey dotted line) anticipated in 2033 is shown. It is clear that the errors of the inter-satellite \acrshort{LRI} dominate over the \acrshort{SGRS} noise in the high-frequency domain.

The colored noise of the modeled \acrshort{SGRS} in the low-frequency domain is mainly caused by thermal instability noise. Therefore, the generated noise time-series includes a low-frequency component ($\sim\qty{10^{-10}}{m/s^2\sqrt{Hz}}$ at $\qty{10^{-4}}{\Hz}$) that causes the drift. These noise time-series were used in the generation of the synthesized noisy observations of the \acrshort{GFR} (see Fig. \ref{fig: closed_loop_procedure}). 

\subsection{Functional model, stochastic modeling and assumptions in GFR simulation}\label{sec_GFR_assumptions}
\acrshort{GFR} simulations were conducted to assess the performance differences between the GRACE-FO satellite shape and its modified counterparts. Following the framework outlined in Fig. \ref{fig: closed_loop_procedure}, synthetic noisy observations were generated for various decaying orbits, incorporating multiple error sources. The simulations considered three accuracy levels of the modeled \acrshort{SGRS} optical accelerometer, accounting for the effects of different satellite shapes, as shown in Fig. \ref{fig: ASD_SGRS_comparison}, as part of the stochastic error component. Additionally, errors arising from the inter-satellite laser ranging instrument were included in the analysis.

The approximate acceleration-based solution approach \citep{Weigelt.2017} is implemented in the \acrshort{QACC} software. In this method, observations are first collected along the orbit at specific measurement positions and subsequently linked to the gradient of the gravitational potential ($\nabla V$). The functional model for \acrshort{ll-SST} observations implemented in the \acrshort{QACC} toolbox is expressed as follows:

\begin{equation}
    \ddot{\vec{\rho}} = \nabla \vec{V}_{AB} \cdot \vec{e}_{AB}^a + \dot{\vec{x}}_{AB} \cdot \dot{\vec{e}}_{AB}^a\text{.}
    \label{eq: QACC_functional_model}
\end{equation}

Here, $\ddot{\vec{\rho}}$ represents the range acceleration, $\nabla \vec{V}_{AB}$ denotes the gradient of the gravitational potential, and $\dot{\vec{e}}_{AB}^a$ is the time derivative of the unit vector along the \acrshort{LoS} between the two satellites. In the \acrshort{QACC} software, the second term of equation \eqref{eq: QACC_functional_model} is neglected \citep{Knabe.2023}. This omission of the so-called centrifugal term removes the orbit-dependent contribution, masks the influence of satellite shapes, and results in an optimistic scenario. The system of linear equations is then solved using least-squares adjustment to determine the spherical harmonic coefficients of the gravity field. According to \citep{Niemeier.2008} the estimated solution $\vec{\hat{x}}$ that minimizes the sum of squares of the weighted residuals in LS adjustment, is given by:

\begin{equation}
    \hat{\vec{x}}=\left( \vec{A}^T\vec{P}\vec{A}\right)^{-1}\vec{A}^T\vec{P}\vec{l}=\vec{N}^{-1}\vec{w}\text{,}
    \label{eq: est_solution}
\end{equation}

where $\vec{A}$ the design matrix, $\vec{P}=\vec{\Sigma_{ll}}^{-1}$ the weight matrix (obtained from stochastic modeling), $\vec{l}$ the observation vector, $\vec{N}$ the normal matrix and $\vec{w}=\vec{A}^T\vec{P}\vec{l}$.

An empirical stochastic modeling approach, involving iterative optimization of the weights, was employed in the gravity field recovery routine for multiple purposes, for example: to down-weight and de-correlate the synthesized colored noise observations \citep{Knabe.2023}, thereby enhancing the signal-to-noise ratio \citep{Luthcke.2013} and improving the estimation of the spherical harmonic coefficients, etc. Initially, the normal matrix $\vec{N}$ is constructed using a unit weight matrix $\vec{P}$, which corresponds to white noise. Afterwards, the full variance-covariance matrix of the measurements $\vec{\Sigma_{ll}}$, is constructed from a biased estimation of the auto-covariance vector $\vec{r}$ and iteratively computed from the post-fit residuals $\vec{\hat{v}}_n$. According to \citep{Koch.2010}, the auto-covariance element $r_i$ is estimated as

\begin{equation}
    r_i = \frac{1}{N}\sum_{n=0}^{N-1-\lvert i\rvert} \hat{v}_n \cdot \hat{v}_{n+i}
    \text{,}
    \label{eq: biased auto-covariance element}
\end{equation}

where $N$ the length of the observations, i.e. 1 month. Finally, assembled variance-covariance matrix of the observations $\vec{\Sigma_{ll}}$  is decomposed by the Cholesky approach.

Empirical stochastic modeling influences the residuals by incorporating all considered noise sources, including inter-satellite \acrshort{LRI} measurement errors, \acrshort{SGRS} noise components, and least-squares adjustment errors. The \acrshort{GFR} results presented in this study were obtained after three iterations of weight optimization, as the solutions showed convergence at this point. The posterior variance $\hat{\sigma}_{0}^2$ of the post-fit residuals quantifies the solution's quality \citep{Wu.2016}:

\begin{equation}
    \hat{\sigma}_{0}^2 = \frac{\vec{\hat{v}}^T\vec{P}\vec{\hat{v}}}{s-r} = \frac{\vec{l}^T\vec{P}\vec{l}-\vec{W}^T\vec{\hat{x}}}{s-r}\text{,}
    \label{eq: posterior_variance}
\end{equation}

where $s$ the number of observations and $r$ the number of parameters.

As outlined in section \ref{sec_software_overview}, the evaluation of results is performed at the level of the residuals by comparing the recovered gravity field model with the reference gravity field. To quantify the differences, unitless spherical harmonic coefficient discrepancies, referred to as true errors, between the reference and recovered gravity field models are multiplied by the Earth's radius, yielding geoid height deviations in meters.

Time-variable background models and the associated aliasing errors represent significant limitations in current satellite gravimetry \citep{Purkhauser.2020}. However, these factors were not considered in this study, as the primary focus is on comparing the performance of the `standard' GRACE-FO satellite shape with the modified configurations, which incorporate the novel \acrshort{SGRS} optical accelerometer. This advanced inertial sensor, along with the anticipated low error levels of the inter-satellite laser range interferometer, has noise characteristics well below the order of magnitude of time-variable background model errors. 

Aliasing effects, which stem from instrument noise and inaccuracies in background models, can significantly impact the accuracy of satellite gravimetry solutions. Improving both of these factors is crucial for achieving more precise results. Time-variable background models are continuously being refined, as demonstrated by advancements such as the release of the Atmosphere and Ocean De-Aliasing Level-1B product RL07 \citep{Shihora.2022} and the recently available AOe07 \citep{Shihora.2024}. Furthermore, the upcoming Mass-change And Geosciences International Constellation (MAGIC) mission, which consists of two satellite pairs, is expected to greatly expand the range of observable mass-change phenomena and resolve significantly smaller spatial scales compared to a near-polar standalone satellite pair \citep{Daras.2023}.

\subsection{Error Analysis}\label{error_evaluation}

Various error sources and underlying assumptions influence the accuracy of the results throughout each simulation phase, including orbital dynamics, SGRS modeling, and gravity field recovery. These factors can introduce deviations that need to be carefully considered. Table \ref{tab:Major Error Sources} provides a summary of the major error sources identified in the simulation procedure.

\newcommand{\tabvspace}{\vspace{2.5pt}}
\begin{table}[!htbp]
    \centering
    \caption{List of major error sources in the simulation procedure.}
    \begin{tabular}{c|c|l} 
        \hline
        \textbf{Software} & \textbf{Type of error} & \textbf{Description} \\ \Xhline{4\arrayrulewidth}
        \multirow{3}{*}{\makecell[c]{\textbf{Orbital dynamics}\\ \textbf{(XHPS)}}} 
        & Integration Errors & \pbox{6cm}{\tabvspace Accumulation of small errors during numerical integration due to finite step size, leading to divergence\tabvspace} \\ \hhline{~--}
        & Numerical Errors & \pbox{6cm}{\tabvspace Truncation errors and round-off errors arise from finite precision of floating-point arithmetic and the omission of higher-order terms\tabvspace} \\ \hhline{~--}
        & Modelling Errors & \pbox{6cm}{\tabvspace Mathematical description of disturbance forces use simplified assumptions\tabvspace} \\ \hline

        \multirow{6}{*}{\makecell[c]{\textbf{SGRS modeling}\\ \textbf{(ACME)}}}
        & Numerical Errors & \pbox{6cm}{\tabvspace Errors from the LPSD (Logarithmic frequency axis Power Spectral Density) algorithm in obtaining the PSD of the instruments from the time domain signals.\tabvspace} \\ \hhline{~--}
        & Modeling Errors & \pbox{6cm}{\tabvspace Assumptions in reference voltage stability may be too optimistic\tabvspace} \\ \hhline{~--}
        & Processing Errors & \pbox{6cm}{\tabvspace Noise budget estimation in the frequency domain using transfer functions may introduce numerical inaccuracies.\tabvspace} \\ \hhline{~--}
        & Measurement Errors & \pbox{6cm}{\tabvspace Sensor imperfections, e.g., thermal noise, capacitive sensing noise, and actuation noise in the SGRS model.\tabvspace} \\ \hhline{~--}
        & Instrumental Errors & \pbox{6cm}{\tabvspace Electrostatic and optical sensor limitations, such as laser interferometric readout accuracy or voltage fluctuations.\tabvspace} \\ \hhline{~--}
        & Alias Errors & \pbox{6cm}{\tabvspace Potential spectral leakage and aliasing due to discrete sampling and windowing effects in the spectral analysis.\tabvspace} \\ \hline

        \multirow{5}{*}{\makecell[c]{\textbf{\acrshort{GFR}}\\ \textbf{(QACC)}}} 
        & \makecell{Approximation Errors} & \pbox{6cm}{%
        \tabvspace
        - The omission of the centrifugal term in the functional model eliminates certain effects and leads to an optimistic scenario.\\
        - The gradient of the gravitational potential was calculated between the positions of two satellites, assuming perfect orbit determination without any attitude errors.\\
        - No scale factor of ACCs or calibration was considered.\\
        - Time-variable background models and associated errors were not included in GFR.
        \tabvspace
        } \\ \hhline{~--}
        & Numerical Errors & \pbox{6cm}{\tabvspace Finite precision of floating-point arithmetic in Fortran.\tabvspace} \\ \hhline{~--}
        & \makecell{Least-Squares \\ Adjustment Errors} & \pbox{6cm}{\tabvspace The inversion of the normal matrix may lead to an amplification of noise.\tabvspace} \\ \hhline{~--}
        & Stochastic Modeling Errors & \pbox{6cm}{\tabvspace Improper noise assumptions, may lead to non-optimal weighting of observations and biased error propagation.\tabvspace} \\ \hhline{~--}
        & Truncation Errors & \pbox{6cm}{\tabvspace GFR solutions were truncated and computed only up to degree and order 90\tabvspace} \\ \hhline{~--}
        & Validation Errors & \pbox{6cm}{\tabvspace Relevant features of the output signal may be not fully captured within EWH or degree RMS representation, potentially leading to the omission or masking of certain characteristics.\tabvspace} \\ \hline
    \end{tabular}
    \label{tab:Major Error Sources}
\end{table}

\newpage
\section{Results}
\label{sec_Res}
\subsection{Satellite orbit simulation and force calulation results}
\label{sec_Sim_Res}

\begin{figure}[!htbp]
    \centering
    \begin{subfigure}[b]{0.55\textwidth}
    	\includegraphics[width=\linewidth]{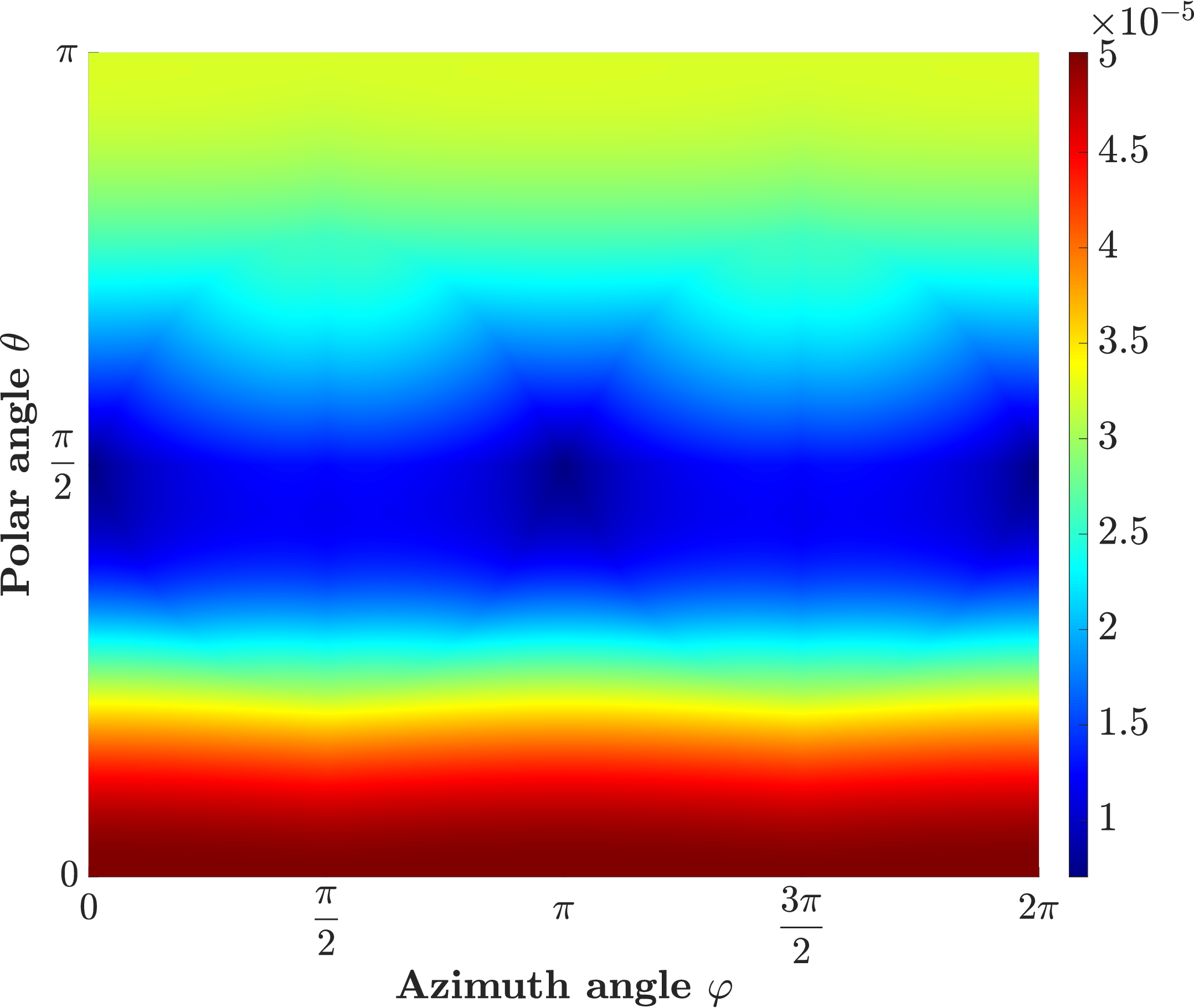}
    	\caption{\acrshort{SRP} force in newton of the standard GRACE-FO satellite shape.}
    	\label{fig: force_GRACE}
    \end{subfigure}
    \hspace{0.2cm}
    \begin{subfigure}[b]{0.4\textwidth}
    	\includegraphics[width=\linewidth]{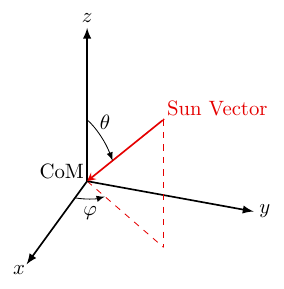}
    	\caption{Satellite center of mass (CoM), satellite reference coordinate system and incident sun vector.}
    	\label{fig: Spherical_Coord_Tht_Phi}
    \end{subfigure}
    \label{fig: force_GRACE_Coord}
    \caption{Precalculated \acrshort{SRP} force for orbital simulation}
\end{figure}

Fig. \ref{fig: force_GRACE} illustrates the variation in \acrshort{SRP} force magnitude as a function of the incident solar rays direction on the satellite geometry calculated by the preprocessing algorithm for the GRACE-FO shape shown in Fig. \ref{fig: fem_GRACE}. Fig. \ref{fig: Spherical_Coord_Tht_Phi} shows the polar angle ($0 - \pi$) and azimuth angle ($0 - 2\pi$) defining the direction of the incoming sun rays relative to the satellite’s reference frame during preprocessing. The color scale encodes the magnitude of the force generated by the incident solar rays. Higher values on the color scale correspond to greater force magnitudes, suggesting stronger interactions between the solar radiation and the satellite surface. This plot effectively captures the anisotropic nature of the force distribution, highlighting how the satellite geometry responds differently to solar radiation depending on the direction of incidence.

\begin{figure}[!hbtp]
	\centering	 
    	\begin{subfigure}[b]{0.45\textwidth}
    		\includegraphics[width=\linewidth]{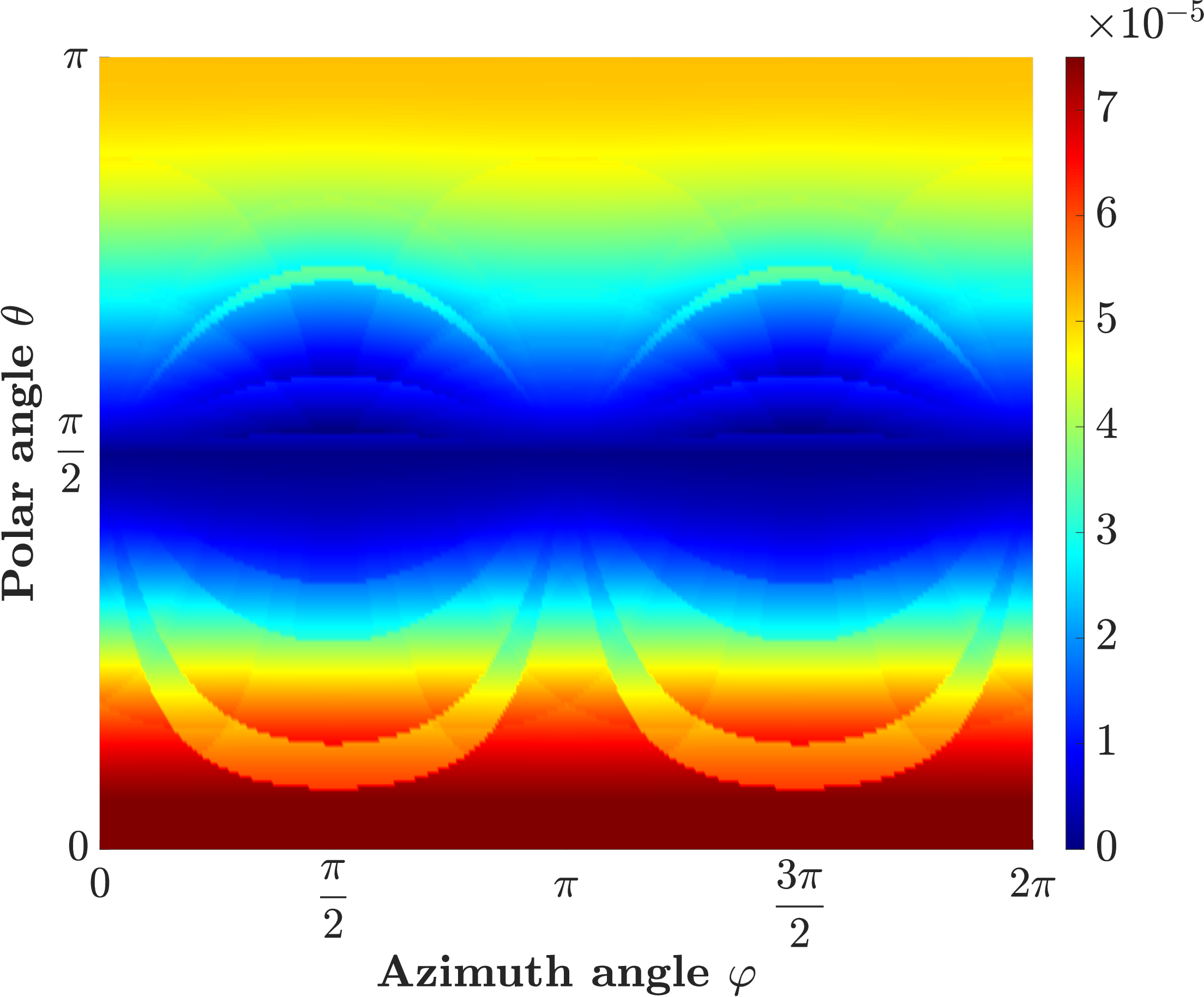}
    		\caption{GRACE-like shape (double solar panel top)}
    		\label{fig: force_GRACE_top}
    	\end{subfigure}
    \hspace{0.2cm}
        \begin{subfigure}[b]{0.45\textwidth}
    		\includegraphics[width=\linewidth]{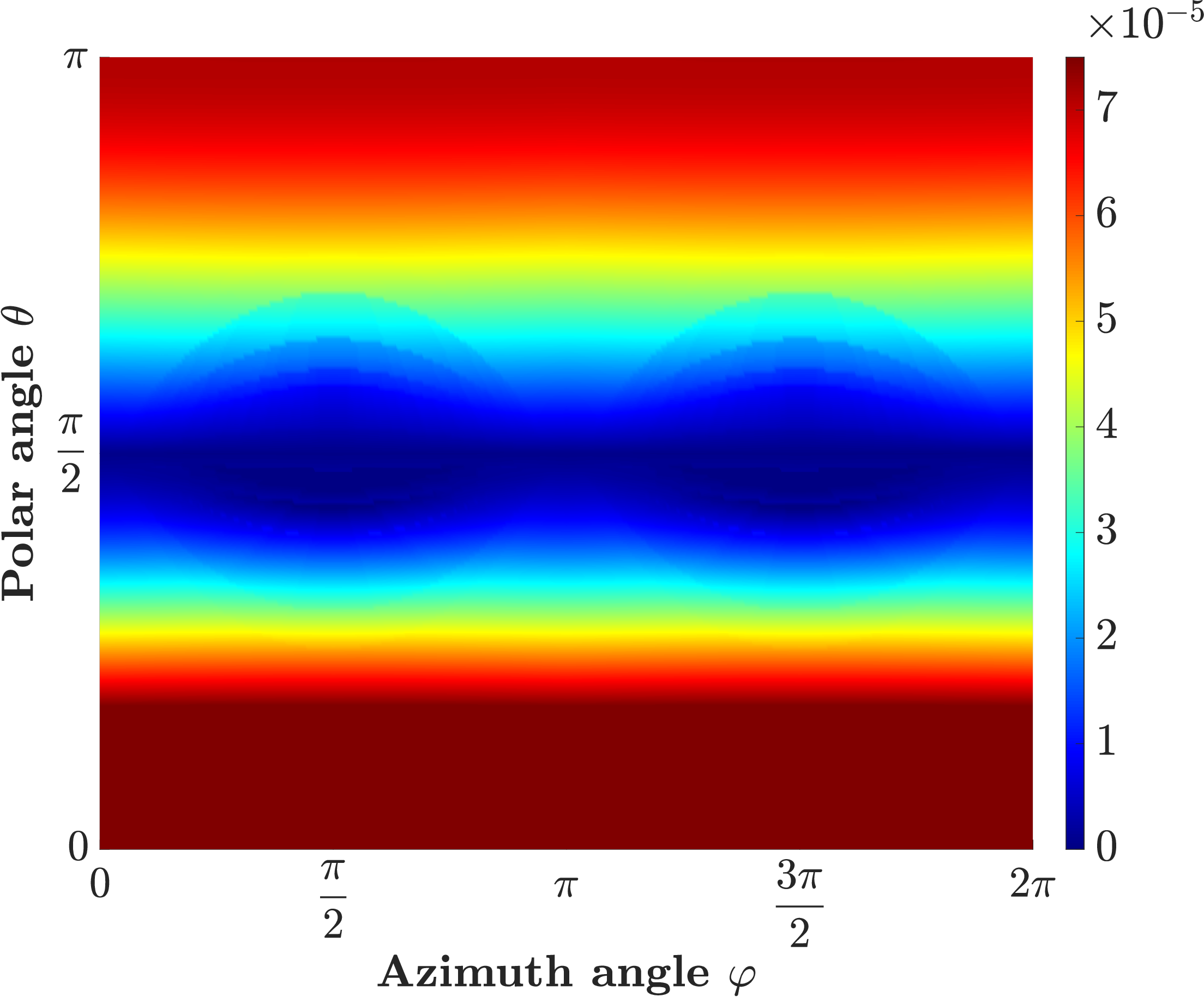}
    		\caption{GRACE-like shape (double solar panel bottom)}
    		\label{fig: force_GRACE_bottom}
    	\end{subfigure}
         \begin{subfigure}[b]{0.45\textwidth}
    		\includegraphics[width=\linewidth]{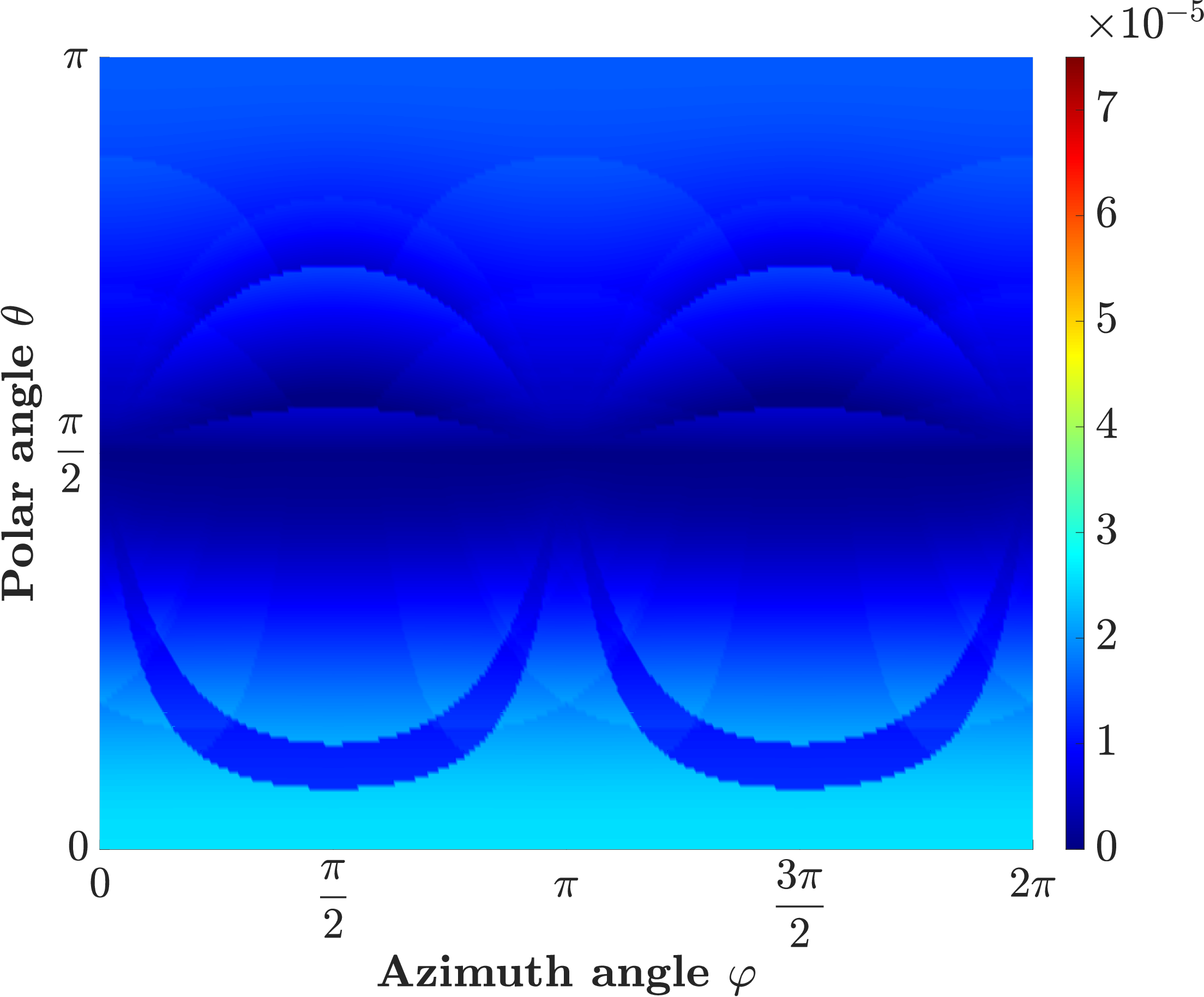}
    		\caption{GRACE-like shape (single solar panel top)}
    		\label{fig: force_GRACE_single_top}
    	\end{subfigure}
    \hspace{0.2cm}
        \begin{subfigure}[b]{0.45\textwidth}
    		\includegraphics[width=\linewidth]{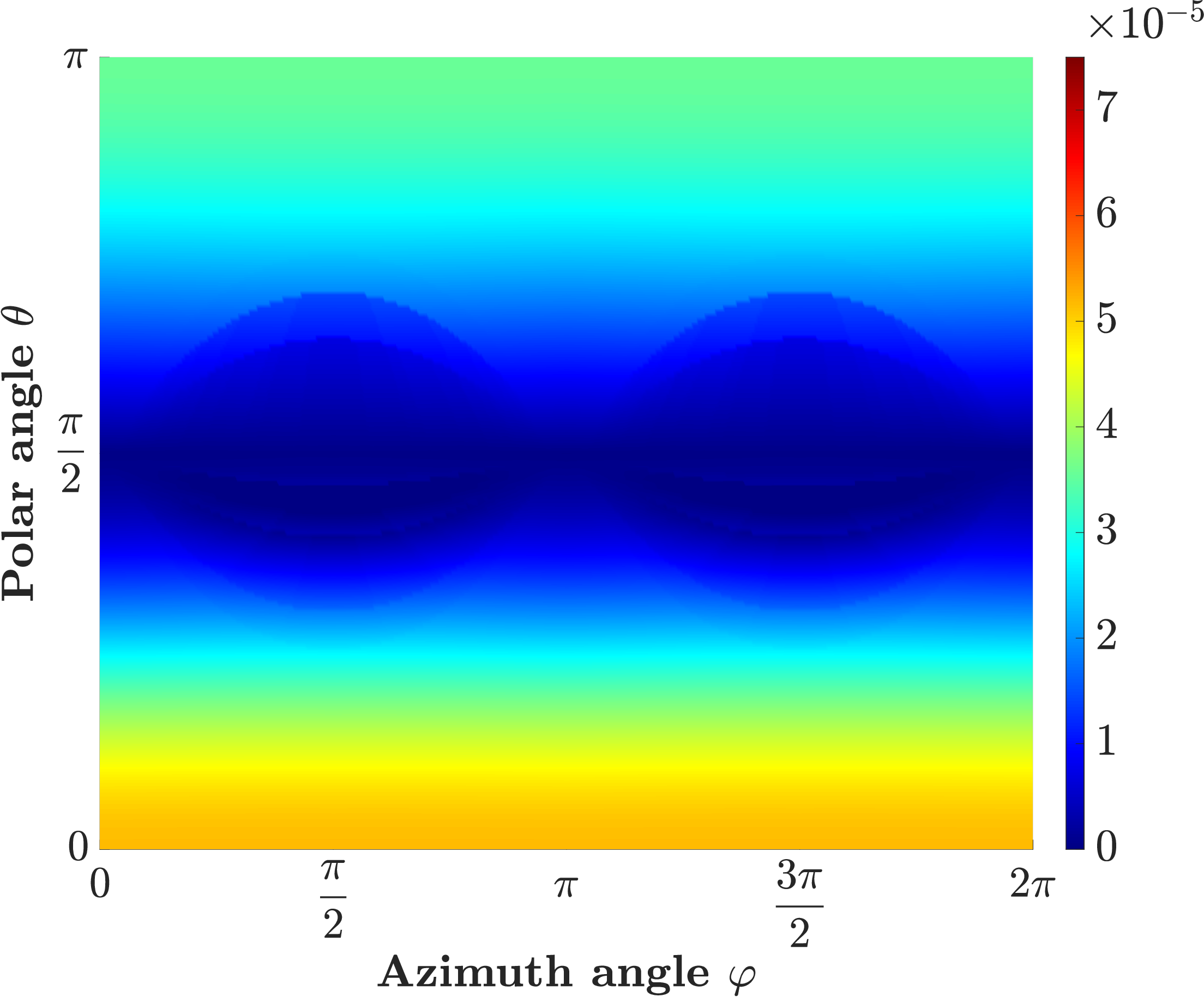}
    		\caption{GRACE-like shape (single solar panel bottom)}
    		\label{fig: force_GRACE_single_bottom}
    	\end{subfigure}	
	\caption{\acrshort{SRP} force difference in newton of the studied satellite shapes to the standard GRACE-FO shape.}
	\label{fig: force_models}
\end{figure}

\begin{figure}[!htbp]
    \centering
    \begin{subfigure}[b]{0.45\textwidth}
    	\includegraphics[width=\linewidth]{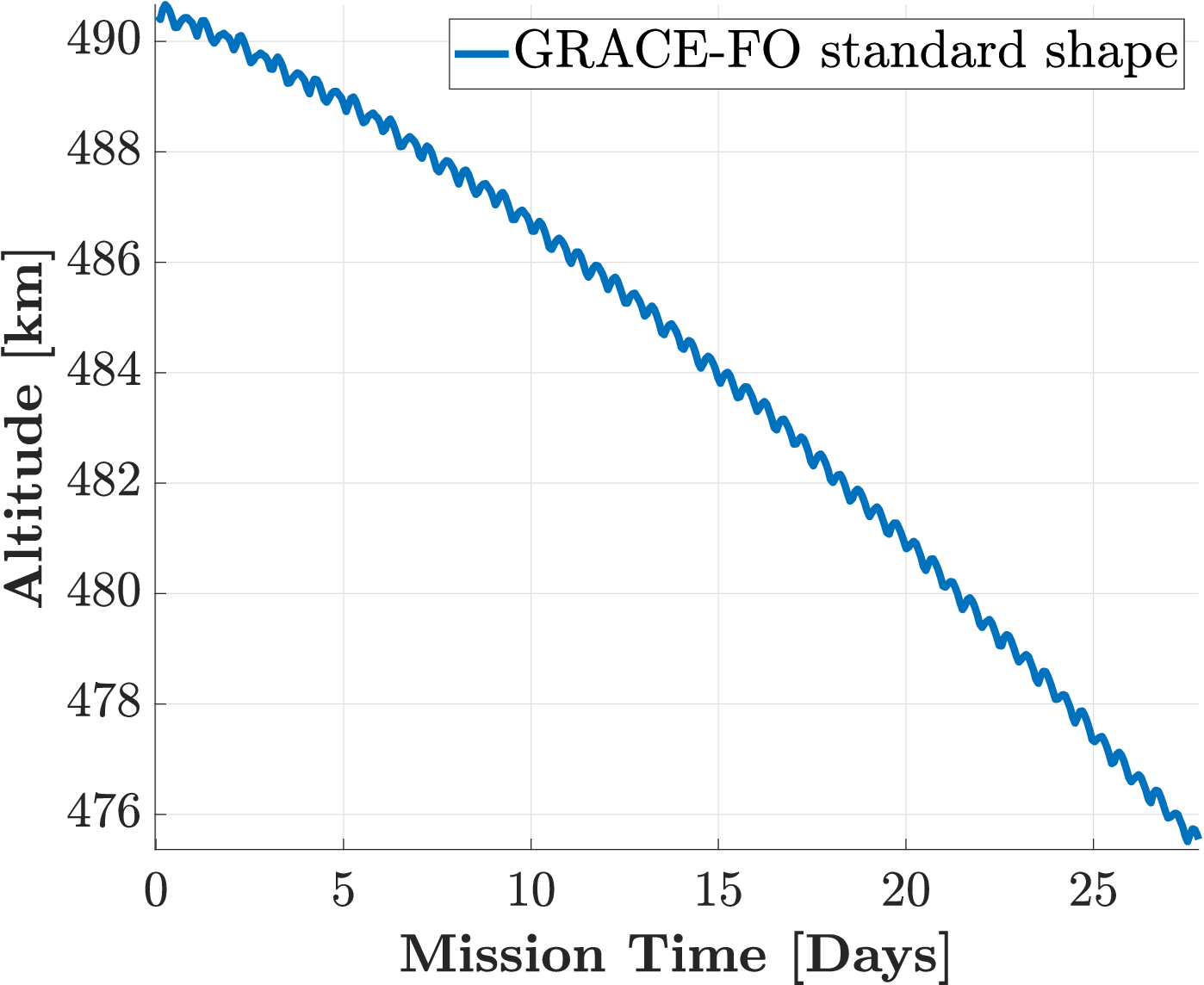}
    	\caption{Simulated decay of the maximum altitude per orbit of the GRACE-FO standard shape case (km).}
    	\label{fig: Alt_GRACE}
    \end{subfigure}
    \hspace{0.2cm}
    \begin{subfigure}[b]{0.45\textwidth}
    	\includegraphics[width=\linewidth]{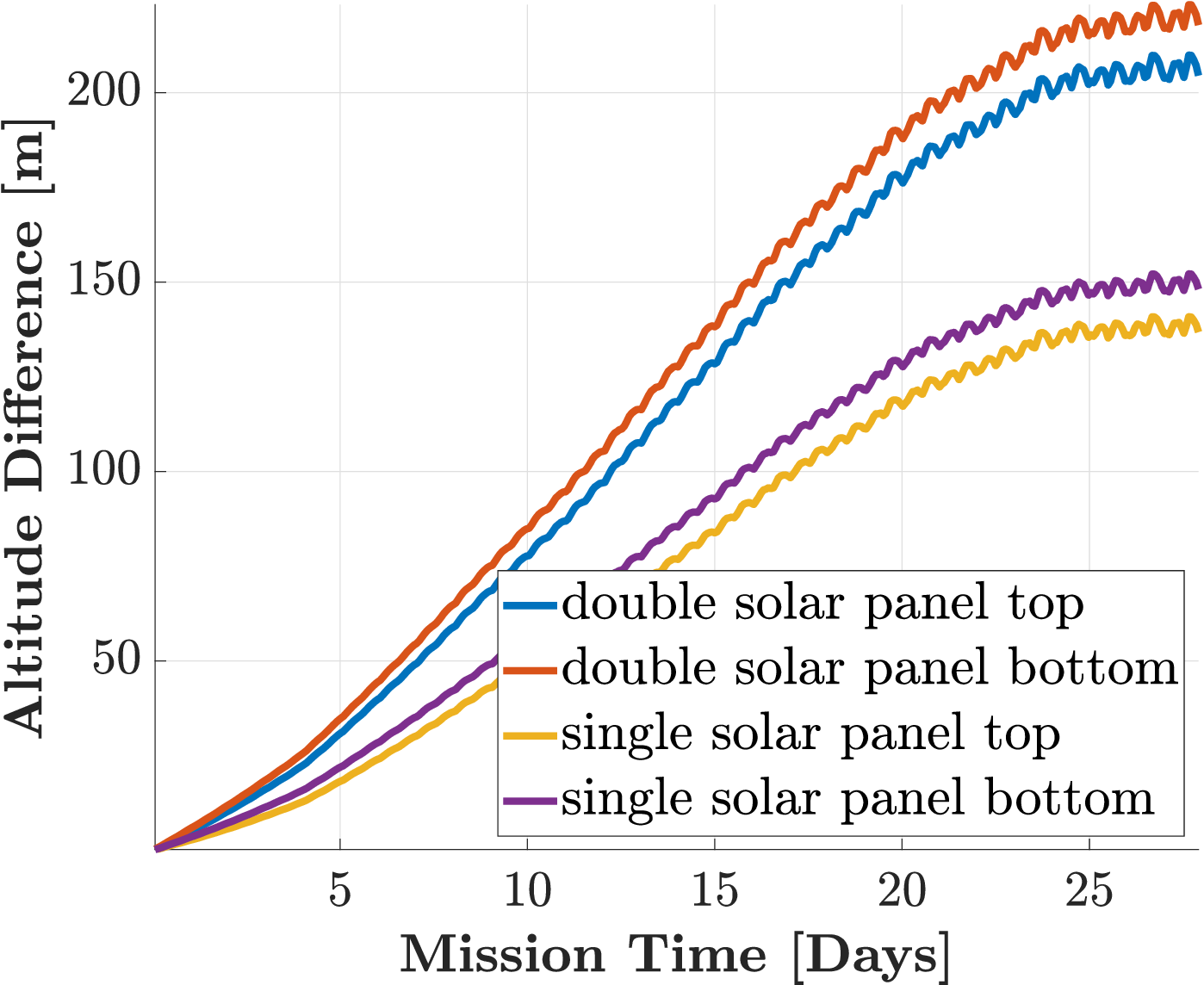}
    	\caption{Difference of the orbit decay of the modified cases ($C_D=3.0$) w.r.t. GRACE-FO standard case (m).}
    	\label{fig: Alt_Diff}
    \end{subfigure}
    \label{fig: orbits_altitude}
    \caption{Absolute altitude of the standard GRACE-FO satellite over a one-month period and differences in altitude for various modified GRACE-like cases relative to the standard configuration. }
\end{figure}

Fig. \ref{fig: force_models} depicts the variation in differences in force magnitude  between each analyzed satellite shape and the GRACE-FO shape. The results indicate that the double solar panel bottom shape exhibits the highest forces, while the single solar panel top shape produces a force profile most similar to that of the GRACE-FO shape. The presence of additional solar panels, as well as the gaps between these panels and the main satellite body, introduces distinct arched like structures in the force profiles, which is due to solar panel gaps allowing light to pass through to the geometry below at certain angles of incidence.

The orbital simulations were conducted under a non-drag-compensated regime, with a drag coefficient of $ c_D = 2.25 $ assigned to the GRACE-FO configuration, and $ c_D = 2.65\text{, }3.0 \text{ and } 4.5 $ for the modified satellite shapes. The orbital decay of the GRACE-FO satellite over a one-month period is shown in Fig. \ref{fig: Alt_GRACE}, represented by the maximum orbital altitude recorded for each orbit. Fig. \ref{fig: Alt_Diff} illustrates the differences in orbital decay between each modified satellite shape (with $ c_D = 3.0$) and the GRACE-FO configuration. The results show that satellite shapes with increased reference areas, and consequently higher drag forces, experience faster orbital decay compared to the GRACE-FO case.

\begin{figure}[!hbtp]
    \centering
    \includegraphics[width=0.6\linewidth]{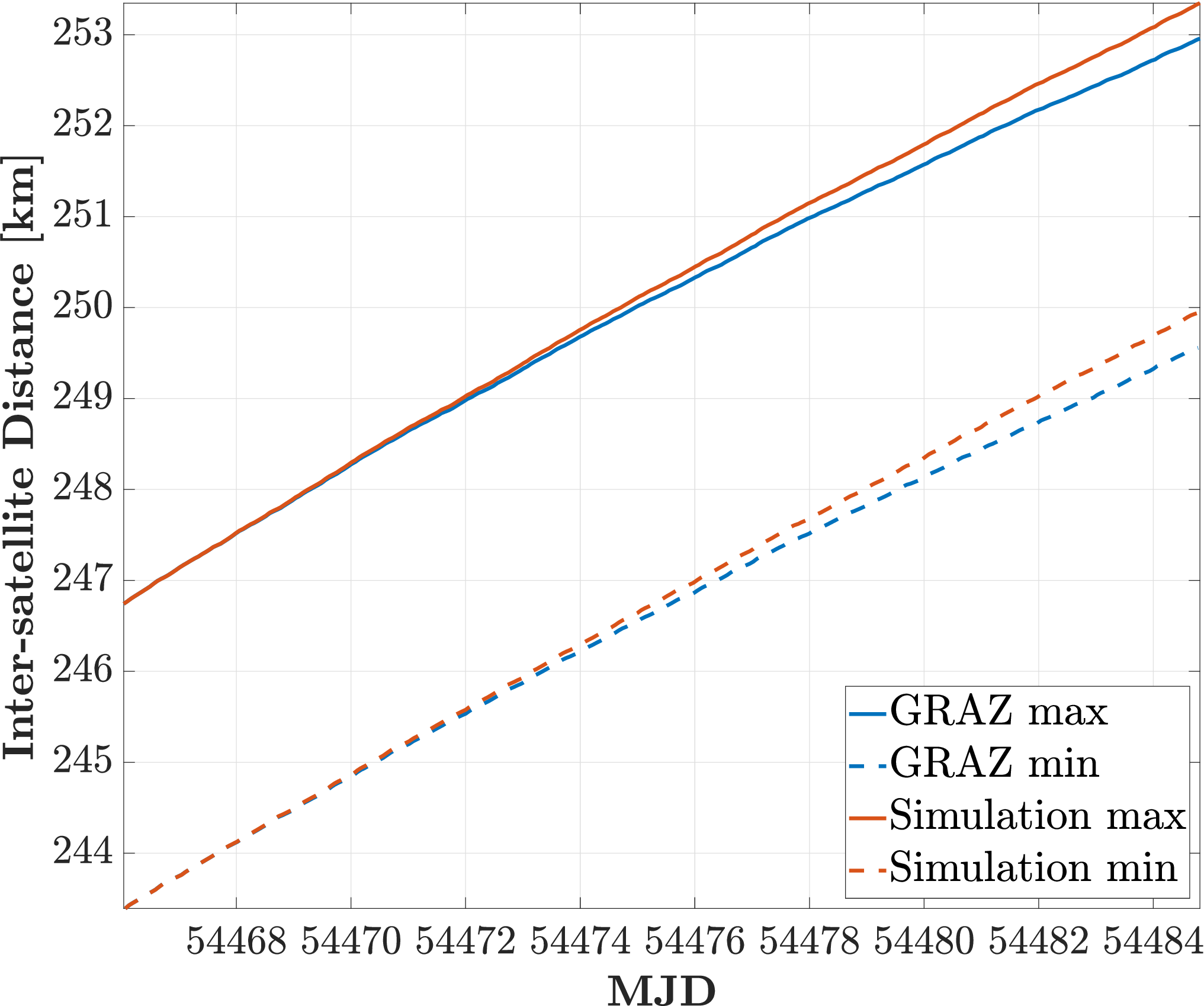}
    \caption{Comparison of the GRACE-FO inter-satellite distance maximum and minimum values obtained from TU Graz \citep{MayerGurr.2021} and simulated in XHPS within this study.}
	\label{fig: InterSatDis}
\end{figure}

Fig. \ref{fig: InterSatDis} illustrates the maximum and minimum values of the inter-satellite distance, defined as the separation between the two satellites during the simulation. The figure also includes the maximum and minimum values of the inter-satellite distance derived from the reduced dynamic orbit data provided by the Technical the University of Graz \citep{MayerGurr.2021}. Initially, the simulated inter-satellite distance aligns closely with the Graz-derived values. However, as the simulation progresses, slight deviations emerge. These discrepancies can be attributed to unmodeled disturbances, modeling inaccuracies, and numerical errors (see table \ref{tab:Major Error Sources}).

\subsection{GFR simulation results}
\label{sec_GFR_Res}

This section presents the retrieved gravity field models from \acrshort{ll-SST} type of missions with the different satellite shapes, diverse performance of the \acrshort{SGRS} optical accelerometers and various air drag coefficients, respectively.

\begin{figure}[!htbp]
    \centering
    \noindent\includegraphics[width=0.75\textwidth]{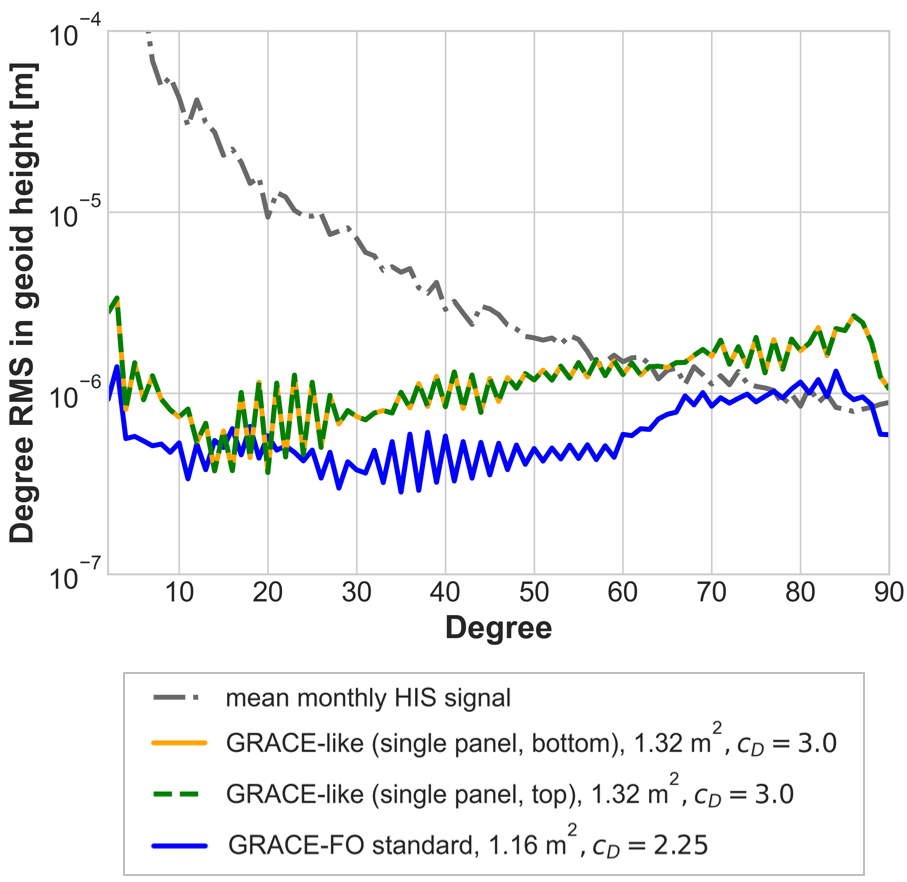}
    \caption{Degree \acrshort{RMS} of the spherical harmonic coefficient differences (true errors) between recovered and EIGEN-6C4 gravity field models in geoid height (m) using the GRACE-FO standard shape (blue curve) and two modified cases with single solar arrays (orange and green lines).}
    \label{fig: GFR_standard_vs_single_panel}
\end{figure}

Fig. \ref{fig: GFR_standard_vs_single_panel} shows a degree \acrfull{RMS} of the spherical harmonic coefficient differences between recovered and reference gravity fields (true errors) for the case of the `standard' GRACE-FO trapezoidal prism shape (blue curve) and of two modified shapes (orange and green curves) with a single solar panel that is mounted either on bottom or top of satellite body. Note that in the `standard' GRACE-FO case a simulated \acrshort{SGRS} optical accelerometer with \acrshort{LRI} 2033 was also considered. Using identical inertial and inter-satellite range sensors enables a fair comparison of different scenarios, highlighting performance variations due to satellite shape differences. In the corresponding orbit simulations, a drag coefficient $c_D=2.25$ was taken for the standard shape and $c_D=3.0$ for the modified ones in order to incorporate the increased surface area from the solar panels (section \ref{sec_xhps}). Relevant cross-sectional areas of the satellite were also considered in deriving the \acrshort{SGRS} noise time series, as its actuation noise depends on the effective area (section \ref{sec_ACC_model}) and depicted in Fig. \ref{fig: ASD_SGRS_comparison}. Also, the grey dashed-dotted line represents the mean monthly time-variable signal from the terrestrial hydrosphere (H), continental ice sheets and mountain glaciers (I) as well as solid Earth deformations (S), or shortly HIS \citep{Dobslaw2015}. Although time-variable background models and their associated errors are among the dominant error sources in current satellite gravimetry missions, they were not included in this GFR study. Since the focus here is on evaluating how modifications to GRACE-like satellites impact gravity field recovery, particularly in relation to the changing performance of inertial sensors (accelerometers) with different satellite shapes. The grey dash-dotted curve is presented to illustrate that, in an idealized scenario where temporal aliasing is sufficiently accounted for and high-performance instruments are fully utilized, a time-variable gravity field could potentially be determined up to approximately degree 60 in both modified cases.

Fig. \ref{fig: GFR_standard_vs_single_panel} also shows that the more complex satellite shapes with a single solar panel, mounted either on top or bottom of the satellite body, do not significantly reduce the quality of the science output. The difference between the modified cases and the standard one, comes, in our optimistic study with multiple simplifications and assumptions  that lead to several error sources and uncertainties listed in the Table \ref{tab:Major Error Sources}, mainly from the different level of the actuation the noise of \acrshort{SGRS} w.r.t. cross-sectorial area  that was explained in section \ref{sec_ACC_model}. The convergence of the orange and green curves for the modified cases can be explained by their similar rates of orbital decay (see Fig. \ref{fig: Alt_Diff}) and the consistent performance levels of the modeled \acrshort{SGRS} (see Fig. \ref{fig: ASD_SGRS_comparison}).

\begin{figure}[!htbp]
    \centering
    \noindent\includegraphics[width=0.75\textwidth]{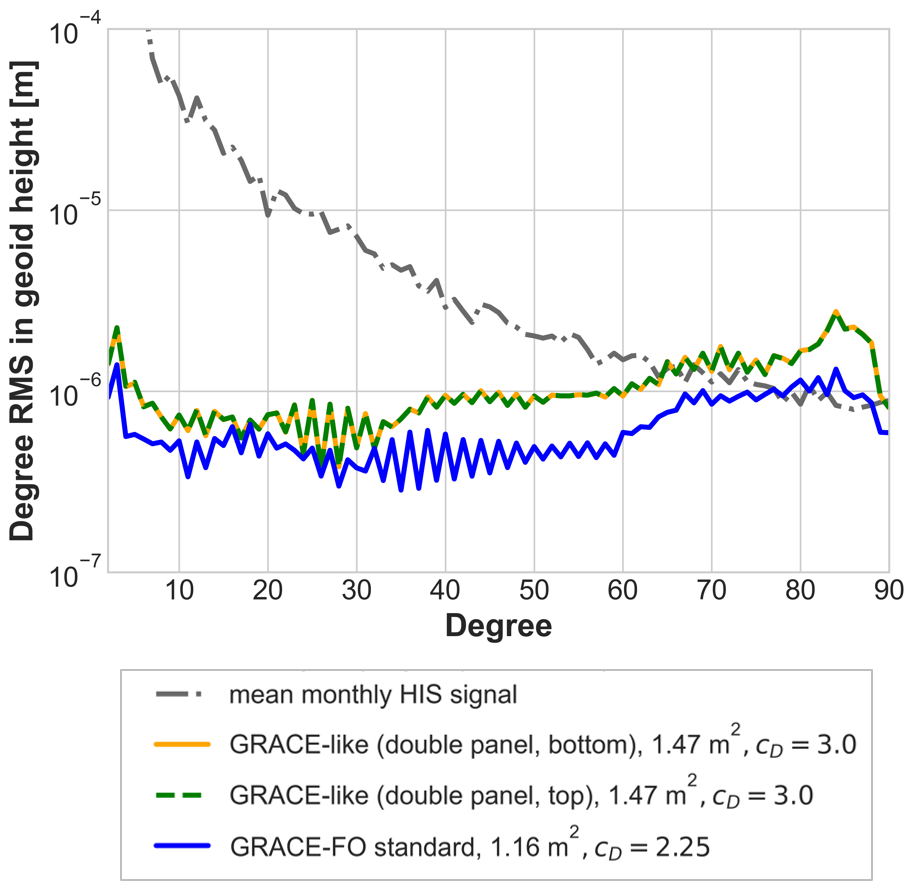}
    \caption{Degree RMS of the spherical harmonic coefficient differences (true errors) between recovered and EIGEN-6C4 gravity field models in geoid height (m) from GRACE-FO standard shape (blue curve) and two modified cases with double solar arrays (orange and green lines).}
    \label{fig: GFR_standard_vs_double_panel}
\end{figure}

Fig. \ref{fig: GFR_standard_vs_double_panel} shows the degree \acrshort{RMS} of the spherical harmonic coefficient differences, plotted in geoid height, for the `standard' GRACE-FO and two modified cases with double solar arrays (cross-sectorial area 1.47 m$^2$) w.r.t. mean monthly HIS signal. Residuals in the spectral domain from the alternative designs are slightly larger than from the standard case. Similar to the previous simulation with a single solar panel, here this difference is mainly due to the different level of the actuation noise of the \acrshort{SGRS} (see Fig. \ref{fig: ASD_SGRS_comparison}). Again, here both advanced scenarios converged due the identical rate of the orbit decay (see Fig. \ref{fig: Alt_Diff}).

Both Figures \ref{fig: GFR_standard_vs_single_panel} and \ref{fig: GFR_standard_vs_double_panel} show that GRACE-like satellites with single or double solar arrays provide slightly lower accuracy in retrieved gravity models compared to the GRACE-FO standard shape. It is important to note that these GFR results reflect the combined influence of orbital dynamics and SGRS performance, both affected by satellite shape. If the optical accelerometer performance were independent of satellite shape, modified satellites—experiencing faster orbit decay (Fig. \ref{fig: Alt_Diff})—would achieve higher gravity field recovery accuracy. However, our carried out GFR simulations indicate that the SGRS performance deteriorates in the mid-frequency domain for modified satellite designs (Fig. \ref{fig: ASD_SGRS_comparison}), outweighing the benefits of stronger orbit decay. In summary, the reduced SGRS performance on modified satellites leads to less accurate GFR solutions, despite their lower orbits compared to the GRACE-FO standard shape.

\begin{figure}[!htbp]
    \centering
    \noindent\includegraphics[width=0.75\textwidth]{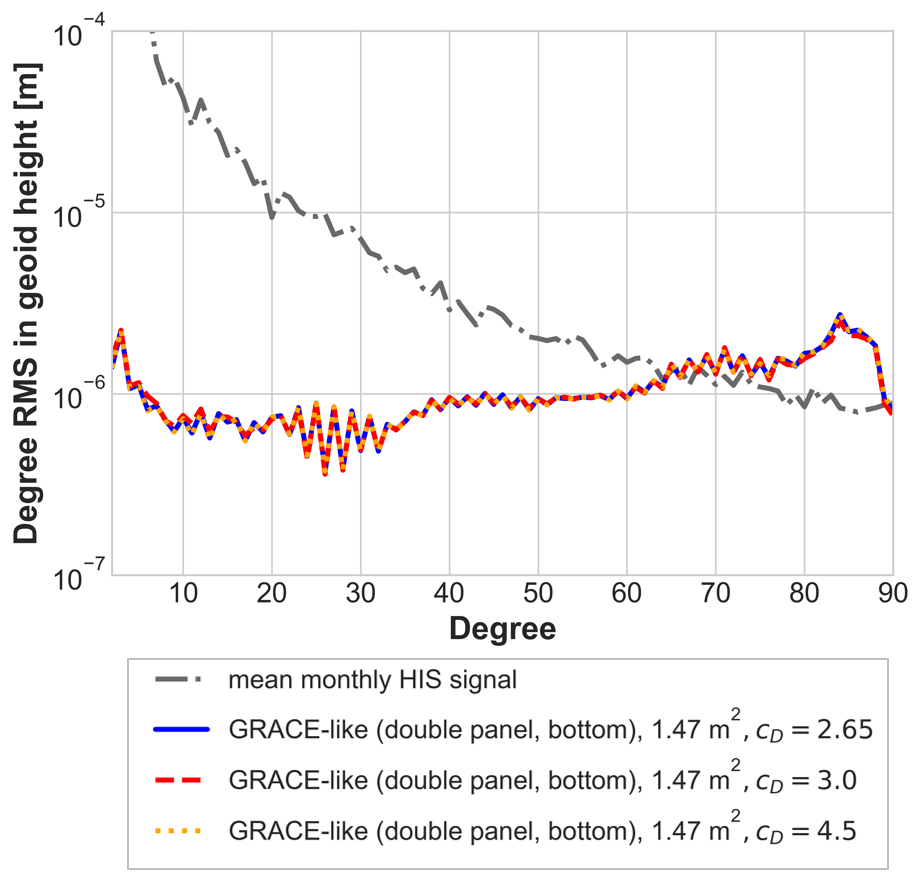}
    \caption{Degree RMS of the spherical harmonic coefficient differences (true errors) between recovered and EIGEN-6C4 gravity field models in geoid height (m) from the GRACE-like shape with a double solar panel and three different drag coefficients.}
    \label{fig: GFR_double_panel_Cds}
\end{figure}

In order to verify whether the observed difference of residuals in the spectral domain between the standard and modified cases with single and double solar arrays, depicted in figures \ref{fig: GFR_standard_vs_single_panel} and \ref{fig: GFR_standard_vs_double_panel}, indeed occur mostly due to the different SGRS performance above $\qty{1}{\mHz}$, additional orbit and GFR simulations have been carried out. Here only a GRACE-like satellite with a double panel mounted on bottom was considered, but with two other air-drag coefficients $c_D=2.65$ and $c_D=4.5$ (representing the critical case, i.e. doubled value of the drag coeffcient for the standard GRACE-FO shape). The same level of accuracy for the inertial sensor and \acrshort{LRI} was considered here. Degree RMS of the spherical harmonic coefficient differences for this GFR simulation are shown in Fig. \ref{fig: GFR_double_panel_Cds}. The GFR of this simulation shows that in spite of different orbit decays due to various drag coefficients, the order of magnitude of the Degree RMS curves in all three cases are the same. Therefore, it can be preliminary concluded that in the non-drag compensated regime, when the orbit decay differs only by a few hundred meters w.r.t. the standard shape scenario (see Fig. \ref{fig: Alt_Diff}), the changing performance of the modeled SGRS w.r.t. the satellite shapes becomes the dominant factor affecting on GFR results.

\section{Discussion and conclusions}\label{ch: Discussion and conclusions}

Future gravimetry missions may require increased power consumption due to drag-compensated regimes, advanced inertial sensors, and non-sun-synchronous orbital configurations. To meet these demands, a larger solar panel area is necessary. Additionally, the mass and volume constraints imposed by launch vehicle limitations make deployable solar panels a practical solution.

In this study closed-loop simulations including detailed orbit dynamics calculations, inertial sensor modeling, and gravity field recovery to evaluate the performance of future \acrshort{ll-SST} gravimetry satellites with four modified shapes have been conducted. Finite element models were developed for the GRACE-FO trapezoidal prism and four GRACE-like satellites, each featuring either single or double solar panels mounted on the top or bottom of the satellite. These models were utilized in orbital simulations performed with \acrshort{XHPS}, under the assumption that the satellite body was completely rigid. Therefore, no vibrations that might negatively impact the \acrshort{SGRS} performance caused by the structural flexibility of the solar arrays were considered and evaluated. The \acrshort{SGRS}, equipped with a laser interferometer \acrshort{TM} readout, was modeled in \acrshort{ACME}, accounting for performance variations due to coupling effects influenced by satellite shape. Finally, \acrshort{GFR} simulations incorporated accelerometer noise time-series, errors in the inter-satellite link, and various orbital scenarios generated in \acrshort{XHPS}.

It is worth noting that eliminating of vibration effects was not the only simplification considered. In the \acrshort{XHPS} satellite dynamics simulator, multiple numerical assumptions were incorporated into the simulations. These include approximations in the modeling of drag, \acrfull{SRP}, orbit propagation, and numerical integration, as well as the exclusion of propellant consumption and orbital maneuvering effects. Similarly, in the \acrshort{GFR} software \acrshort{QACC}, the centrifugal term in the functional model was omitted, along with the effects of time-variable background models. Additionally, in the modeling of the \acrshort{SGRS} noise budget within \acrshort{ACME}, further assumptions were made, such as treating the system as a single degree of freedom and considering only air drag as non-gravitational perturbation.

Neverthless, this simplified study provides a good preliminary assessment of the impact of different satellite shapes in a GRACE-like \acrfull{ll-SST} configuration on gravity field recovery. The results show that the residuals between the recovered and reference gravity field models obtained for the five satellite shapes are of a similar order of magnitude. However, as demonstrated in the orbital simulations, satellites equipped with extended solar panels exhibit a faster orbit decay compared to the standard GRACE-FO design. Specifically, a larger decay of \qty{200}{\m} per month was observed for shapes with double solar panels compared to the standard GRACE-FO design, while the single-panel shapes exhibited a decay of approximately \qty{150}{\m}. These values would increase significantly under conditions of high solar activity or less accurate satellite attitude control systems. 

Moreover, the increased surface area associated with extended solar panels leads to larger disturbance torques acting on the satellite, resulting in greater activity of the attitude control system to maintain attitude stability. Over time, higher disturbance torques can cause reaction wheels to reach saturation more frequently, requiring more frequent momentum desaturation. This may result in additional thruster activations, leading to higher fuel consumption.

An essential consideration emerging from this study is the trade-off presented by the modified satellite shapes: while the extended panels may support higher power requirements for future instruments, they also introduce challenges related to mission lifetime and the accuracy of the retrieved gravity models. This trade-off underscores the importance of a holistic approach to satellite design that balances these competing demands.

Another critical aspect that has to be taken into account, when utilizing the extended solar panels, is the field of view of the star cameras. They are located on the skewed sides of the satellite body and might be obscured by the extra solar panels. Therefore, certain workarounds are needed to maintain the necessary level of accuracy from the star cameras. One possible solution could be incorporating holes in the solar panels, when considering top mounted solar panels.

 Further studies should focus on addressing the limitations and simplifying assumptions made in this investigation to provide a more comprehensive understanding of the implications of modified satellite shapes on gravity field recovery performance. In principle, each of the considered software tools could be improved by considering additional parameters or enhanced models and routines. Incorporating time-variable background models and associated aliasing effects would allow for a more realistic simulation of operational conditions. Additionally, refining the modeling of non-gravitational forces by including satellite vibrations and propellant consumption would improve the accuracy of the satellite dynamics simulations. Enhanced thermal noise modeling and a broader exploration of different inertial sensor configurations, including advanced hybrid or quantum-based sensors, could also be pursued. Finally, the analysis should be extended to include variations in orbit scenarios, such as high solar activity periods and alternative orbital inclinations, to better evaluate the trade-offs between mission duration, attitude stability, and gravity field recovery accuracy.

\bmhead{Funding}
The authors acknowledge funding by Deutsche Forschungsgemeinschaft (DFG) - TerraQ (Project-ID 434617780 - SFB 1464)

\bmhead{Conflict of interest}
The authors declare that they have no known competing financial interests or personal relationships that could have appeared to influence the work reported in this paper.

\bmhead{Data availability}
The datasets generated during and/or analysed during the current study are available from the corresponding author on reasonable request.

\bmhead{Author Contributions}
A. Le., A. Ku., A. Re., A. Kn., M. Sc. and M. Li. contributed to conceptualization; A. Le., A. Ku., A. Re. and M. We. provided methodology; A. Le., A. Ku. and A. Re. performed formal analysis, simulation and investigation; A. Le., A. Ku. and A. Re. performed writing-original draft preparation; A. Le., A. Ku., A. Re., A. Kn., M. Sc., M. Li., V. Mü., M. We., J. Mü. performed writing-review and editing.

\bibliography{bibliography}

\end{document}